\documentclass[prd,nofootinbib,floats,superscriptaddress,eqsecnum,tightenlines,11pt]{revtex4}
\pdfoutput=1
\usepackage{hyperref}
\usepackage{graphicx}
\usepackage{amsmath,amssymb,amsfonts,amsthm,latexsym,stmaryrd}
\usepackage{marginnote}
\usepackage{color}
\usepackage{enumerate}
\usepackage{float}

\newcommand{\no}{\noindent}
\newcommand{\nb}{\nonumber}
\def\beq{\begin{equation}}
\def\eeq{\end{equation}}
\newcommand{\bea}{\begin{eqnarray}}
\newcommand{\eea}{\end{eqnarray}}
\def\bi{\begin{itemize}}
\def\ei{\end{itemize}}
\def\ba{\begin{array}}
\def\ea{\end{array}}
\def\bfig{\begin{figure}}
\def\efig{\end{figure}}

\def\tA{\hat A} 
\def\B{{\cal B}}
\def\K{{\cal K}}

\def\s{\sigma} 
\def\m{\sigma}

\def\aa{a_1}
\def\ab{a_2}
\def\ac{a_3}
\def\ad{a_4}
\def\ae{a_5}
\def\ba{\beta_1}

\def\b{b}

\def\An{A_*}
\def\dotAn{\dot A_*}
\def\A{{\cal A}}
\def\V{{\cal V}}

\def\l{\lambda}
\def\tC{\tilde C}

\newcommand\Ost{Ostrogradsky }
\newcommand\Ga{\Gamma}
\newcommand\Om{\Omega}

\def\l{\ell}

\begin{document}

\title{
Degenerate higher order scalar-tensor theories beyond Horndeski \\
up to cubic order}
\author{J. Ben Achour}
\affiliation{Center for Field Theory and Particle Physics, Fudan University, 20433 Shanghai, China}
\author{M. Crisostomi}
\affiliation{Institute of Cosmology and Gravitation, University of Portsmouth, Portsmouth, PO1 3FX, UK}
\author{K. Koyama}
\affiliation{Institute of Cosmology and Gravitation, University of Portsmouth, Portsmouth, PO1 3FX, UK}
\author{D.  Langlois}
\affiliation{Laboratoire APC -- Astroparticule et Cosmologie, Universit\'e Paris Diderot Paris 7, 75013 Paris, France}
\author{K. Noui}
\affiliation{Laboratoire de Math\'ematiques et Physique Th\'eorique, Universit\'e Fran\c cois Rabelais, Parc de Grandmont, 37200 Tours, France}
\affiliation{Laboratoire APC -- Astroparticule et Cosmologie, Universit\'e Paris Diderot Paris 7, 75013 Paris, France}
\author{G. Tasinato}
\affiliation{Department of Physics, Swansea University, Swansea, SA2 8PP, UK \\} 

\begin{abstract}
\noindent
We present all scalar-tensor Lagrangians that are cubic in second derivatives of a scalar field, and that are degenerate, hence avoiding Ostrogradsky instabilities. Thanks to the existence of constraints, they propagate no more than three degrees of freedom, despite having higher order equations of motion. We also determine the viable combinations of previously identified quadratic degenerate Lagrangians and the newly established cubic ones.    
Finally, we study whether the new theories are connected to known scalar-tensor theories such as Horndeski and beyond Horndeski, through conformal or disformal transformations.
\end{abstract}

\maketitle

\section{Introduction}
\no General Relativity (GR) is the unique consistent classical theory for a massless, self-interacting  spin two field in four dimensional spacetime \cite{Weinberg:1972kfs}. It describes  accurately gravitational phenomena spanning  a large  interval of scales, from short distances  probed by table top experiments, to large  distances probed by astronomy and astrophysics \cite{Will:2014kxa}. By including a positive cosmological constant term to the Einstein-Hilbert action, GR can also describe the  current acceleration of the universe, but only if one is willing  to accept the enormous fine tuning  that observations require on the value of the cosmological constant  \cite{Weinberg:1988cp}.
Attempts to avoid  such fine tuning  motivate the study of gravitational theories  more general than GR,  the simplest option being  scalar-tensor theories of gravity (see e.g. \cite{Fujii:2003pa} for a review). Theories that involve derivative scalar interactions, in the family of Galileons \cite{Nicolis:2008in}, are characterised by  interesting screening effects, as for example the Vainsthein mechanism \cite{Vainshtein:1972sx}, which are able to reduce the strength of the scalar fifth force to a value compatible with present constraints on deviations from GR. 

Intriguingly,  although the subject has been studied for many decades by now, we still do not know 
the structure of the  {\it most general} consistent scalar-tensor theory, i.e. a theory describing a scalar interacting with a spin-2 tensor field in four dimensions. Horndenski \cite{horndeski} analysed the most general actions for scalar-tensor theories which lead to second order equations of motion (EOMs),
and avoid Ostrogradsky instabilities  \cite{Ostrogradski}.
In four dimensional spacetime, this condition allows one to consider actions which contain at most three powers of second derivatives
of the scalar field.  However, as realised only recently, there also exist viable theories   ``beyond Horndeski''~\cite{Zumalacarregui:2013pma,Gleyzes:2014dya,Gleyzes:2014qga}, which do not suffer from the 
Ostrogradsky instability even though the corresponding  Euler-Lagrange equations are higher order. 
Such theories
 have interesting  consequences for cosmology and astrophysics. 
 In particular, they lead to a breaking of the Vainshtein mechanism {\it inside} matter,  which can modify the structure of  nonrelativistic stars ~\cite{Kobayashi:2014ida,Koyama:2015oma,Saito:2015fza,Sakstein:2015zoa,Sakstein:2015aac,Jain:2015edg}, as well as that of   relativistic ones~\cite{Babichev:2016jom}. 

The aim of the present paper is to determine  the maximal generalization of  Horndenski theories in four dimensions,  by which we mean all scalar-tensor theories that contain at most three powers of second derivatives of the scalar field, and that  propagate at most three degrees of freedom. 

\smallskip

As demonstrated in \cite{LN1}, a systematic way to identify scalar-tensor theories  that contain at most  three degrees of freedom, i.e. without Ostrogradsky ghost,  is to consider  Lagrangians that are degenerate, i.e. whose Hessian matrix -- obtained by taking the second derivatives of the Lagrangian with respect to velocities -- is degenerate. For scalar-tensor theories, such a degeneracy can depend on the specific coupling between the metric and the scalar field. From the Hamiltonian point of view, the degeneracy of the Lagrangian translates into the existence of constraints on phase space, in addition to the usual Hamiltonian and momentum constraints due to diffeomorphism invariance, and explains why one degree of freedom is eliminated, even  if the equations of motion are higher order. A detailed Hamiltonian analysis confirms the direct link between this degeneracy and the elimination of the Ostrogradsky ghost~\cite{LN2}. 
For Lagrangians depending on the accelerations of {\it several} variables,  the degeneracy of the Lagrangian is not sufficient to eliminate the multiple Ostrogradsky ghosts and extra conditions must be imposed, as shown  in  \cite{Motohashi:2016ftl}  for classical mechanics systems (see also  \cite{Klein:2016aiq} for a slightly different approach, reaching the same conclusion).
The singularity of the Hessian matrix (this time obtained by taking the second derivatives of the Lagrangian with respect to the lapse and shift) finds application also in other contexts like massive gravity:
indeed, it is this condition that provides the tertiary constraint necessary to remove the Boulware-Deser ghost mode \cite{Comelli:2012vz}.

The degeneracy criterium, which provides a powerful and simple method to identify viable theories, was used in \cite{LN1} to  identify all scalar tensor theories whose Lagrangian depends quadratically on second order derivatives of a scalar field. Degenerate higher derivative Lagrangians, later dubbed EST (Extended Scalar Tensor) in \cite{Crisostomi:2016czh}, or  DHOST (Degenerate Higher Order Scalar Tensor) in \cite{Achour:2016rkg}, include  Horndeski theories  as well as their extensions ``beyond Horndeski''. As stressed in \cite{LN1}  and \cite{Crisostomi:2016tcp}, only specific combinations of Horndeski theories and of their extensions beyond Horndeski are (Ostrogradsky) ghost-free. 
Quadratic degenerate theories are further studied in \cite{Crisostomi:2016czh,Achour:2016rkg,deRham:2016wji}, in particular how they change under disformal transformations of the metric.

In the present work, we extend the systematic classification of  degenerate theories to include Lagrangians that possess a cubic dependence  on second order derivatives, so to find the most general extension of Horndenski scalar-tensor theory of gravity. We also allow for non-minimal couplings with gravity 
and show that the {\it only} viable Lagrangian, among all possible ones involving the Riemann tensor contracted with the second derivative of the scalar field, is of the form~$G^{\mu\nu}\nabla_\mu\!\nabla_\nu\phi$.
The class of theories we consider thus encompasses Horndeski Lagrangians and our analysis confirms that all Horndeski theories are degenerate, as expected. We also find new classes of cubic Lagrangians that are degenerate.  In total, we identify seven classes of minimally coupled cubic theories, and two classes of non-minimally coupled cubic theories. We study in which cases it is possible to combine any of these cubic theories with the previously identified  quadratic theories to obtain more general Lagrangians. 
We investigate which cubic theories admit a well-defined Minkowski limit, i.e. when the metric is frozen to its Minkowski value.
We also study whether the new cubic theories are related to known Lagrangians through conformal or disformal transformations.  Technical appendixes contain details of the calculations leading to the results we present in the main text. 
   
\section{Degenerate Scalar-Tensor Theories}

\no Scalar-Tensor theories involving second order derivatives of the scalar field in the action are  generally plagued by an Ostrogradsky instability, unless the Lagrangian is degenerate, i.e. there is a primary constraint that leads to the removal of 
the additional undesired mode\footnote{In this paper we do not perform a full Hamiltonian analysis 
(see Refs.~\cite{LN2} and \cite{Deffayet:2015qwa} for Hamiltonian formulations of beyond Horndeski theories);
however, we expect that, in general,  the primary constraint  is second-class and leads to a secondary constraint that is also second-class, so that both constraints remove one degree of freedom, as shown explicitly in \cite{LN2} for the quadratic case. Note that the primary constraint can also be first-class in some very particular cases.}.
In order to study these theories, it is useful to recast the action into ordinary first 
order form via the introduction of a suitable auxiliary variable.
This can be done by  
replacing all first order derivatives $\nabla_\mu\phi$ by the components of 
a vector field $A_\mu$, as first explained in \cite{LN1}, and 
by imposing the relation 
\bea
A_\mu = \nabla_\mu\phi \,, \label{lagmult}
\eea
using  a Lagrangian multiplier.
Therefore, after introducing the general action we investigate, we will focus on  its kinetic structure by identifying the time derivatives of the fields contained in~$\nabla_{\mu} A_{ \nu}$.  
 
\subsection{Action}
\def\2{{(2)}}
\def\3{{(3)}}
\no In this paper we consider the most general action involving quadratic and cubic powers of the second derivative of the scalar field:
\bea
S[g,\phi] = \int d^4 x \, \sqrt{- g }
\left( f_2 \, R + C_\2^{\mu\nu\rho\sigma} \,  \phi_{\mu\nu} \, \phi_{\rho\sigma}
+ f_3 \, G_{\mu\nu} \phi^{\mu\nu}  +  
C_\3^{\mu\nu\rho\sigma\alpha\beta} \, \phi_{\mu\nu} \, \phi_{\rho\sigma} \, \phi_{\alpha \beta} \right) \label{general action} \,,
\eea
where the functions $f_2$  and $f_3$ depend only on $\phi$ and $X \equiv \nabla_\mu \phi \nabla^\mu \phi$ (we use a mostly plus convention for the spacetime metric). The tensors 
$C_\2$ and $C_\3$ are the most
general tensors constructed with the metric $g_{\mu\nu}$ and the first derivative of the scalar field 
$\phi_\mu \equiv \nabla_\mu \phi$.

As we will see in detail in the next subsection, when written in terms of the auxiliary variable $A_\mu$, each 
second derivative of $\phi$ yields terms linear in velocities. By contrast, the curvature depends  quadratically on the velocities of the metric and one can introduce terms non-minimally coupled to gravity, such as $f_2 \, R$ and  $f_3 \, G_{\mu\nu} \phi^{\mu\nu} $, leading  to second or third powers in velocities respectively. 
A priori, one could also envisage many more terms of this kind  involving the Riemann tensor contracted in various ways. However, as shown in Appendix \ref{Nterms},  the {\it only} viable Lagrangians among all the possible ones with appropriate powers in velocities, turn out to be these two (up to integrations by parts). 

Note that one could also include in our general action \eqref{general action} terms of the form $P(X,\phi)$ or terms depending linearly on $\phi^{\mu\nu}$. We have not included such terms explicitly because they do not modify the degeneracy conditions, but one should keep in mind that they can always be added to the Lagrangians that will be identified in our analysis.

Due to the way the tensors  $C_\2$ and $C_\3$
are contracted in the action, one can always impose, without loss of generality,  the symmetry relations:
\bea
C_\2^{\mu\nu\rho\sigma} =C_\2^{\rho\sigma\mu\nu}=  C_\2^{\nu\mu\rho\sigma } 
\quad \text{and} \quad
C_\3^{\mu\nu\rho\sigma\alpha\beta} = C_\3^{\rho\sigma \mu\nu \alpha\beta} = C_\3^{ \mu\nu \alpha\beta \rho\sigma} =
C_\3^{ \nu\mu \alpha\beta \rho\sigma} \,. \label{sym}
\eea
As a consequence, they can be expressed as 
\bea
C_\2^{\mu\nu\rho\sigma} & = &  \langle \langle  \aa\,  g^{\mu\rho} g^{\nu\sigma} +
\ab \,g^{\mu\nu} g^{\rho\sigma} + \ac\, \phi^\mu\phi^\nu g^{\rho\sigma} 
+   \ad \phi^\mu \phi^\rho g^{\nu\sigma}  +  \ae\, \phi^\mu \phi^\nu \phi^\rho \phi^\sigma \rangle\rangle
\label{C2}\,, \\ 
 C_\3^{\mu\nu\rho\sigma\alpha\beta} & = &\langle\langle \b_1 \, g^{\mu\nu} g^{\rho\sigma} g^{\alpha \beta} + 
 \b_2 \, g^{\mu\nu} g^{\rho \alpha} g^{\sigma \beta} +  \b_3 \, g^{\mu \rho} g^{\nu\alpha} g^{\sigma \beta} 
+  \b_4 \, g^{\mu\nu} g^{\rho\sigma} \phi^\alpha \phi^\beta 
 \nb \\
 &&  +\b_5 \, g^{\mu \nu} g^{\rho\alpha} \phi^\sigma \phi^\beta
+  \b_6 \, g^{\mu\rho} g^{\nu\sigma} \phi^\alpha \phi^\beta 
  +  \b_7 \, g^{\nu\rho} g^{\sigma\alpha} \phi^\mu \phi^\beta
  + \b_8 \, g^{\mu\rho} \phi^\nu \phi^\sigma \phi^\alpha \phi^\beta 
  \nb\\
 && +    \b_9 \, g^{\mu\nu} \phi^\rho \phi^\sigma \phi^\alpha \phi^\beta
+  \b_{10} \, \phi^{\mu}  \phi^{\nu}  \phi^{\rho}  \phi^{\sigma}
  \phi^{\alpha}  \phi^{\beta} \rangle\rangle \,,
  \label{C3}
\eea
where the functions $a$'s  and $ \b$'s depend only on $\phi$ and $X$. 
The notation $\langle\langle \dots\rangle\rangle$ means that these expressions are
symmetrised so to satisfy eq  \eqref{sym}.
Explicitly, we have
\beq
 C_\2^{\mu\nu\rho\sigma} \,  \phi_{\mu\nu} \, \phi_{\rho\sigma}+  
C_\3^{\mu\nu\rho\sigma\alpha\beta} \, \phi_{\mu\nu} \, \phi_{\rho\sigma} \, \phi_{\alpha \beta} =\sum_{i=1}^{5}a_i  L^\2_ i + \sum_{i=1}^{10} b_i L^\3_i \,,
\eeq
where 
\bea
&& L^\2_1 = \phi_{\mu \nu} \phi^{\mu \nu} \,, \qquad
L^\2_2 =(\Box \phi)^2 \,, \qquad
L_3^\2 = (\Box \phi) \phi^{\mu} \phi_{\mu \nu} \phi^{\nu} \,,  \nb \\
&& L^\2_4 =\phi^{\mu} \phi_{\mu \rho} \phi^{\rho \nu} \phi_{\nu} \,, \qquad
L^\2_5= (\phi^{\mu} \phi_{\mu \nu} \phi^{\nu})^2\,;
\eea
and
\bea
&& L^\3_1=  (\Box \phi)^3  \,, \quad
L^\3_2 =  (\Box \phi)\, \phi_{\mu \nu} \phi^{\mu \nu} \,, \quad
L^\3_3= \phi_{\mu \nu}\phi^{\nu \rho} \phi^{\mu}_{\rho} \,,  \nb \\
&& L^\3_4= \left(\Box \phi\right)^2 \phi_{\mu} \phi^{\mu \nu} \phi_{\nu} \,, \quad
L^\3_5 =  \Box \phi\, \phi_{\mu}  \phi^{\mu \nu} \phi_{\nu \rho} \phi^{\rho} \,, \quad
L^\3_6 = \phi_{\mu \nu} \phi^{\mu \nu} \phi_{\rho} \phi^{\rho \s} \phi_{\s} \,, \nb  \\
&& L^\3_7 = \phi_{\mu} \phi^{\mu \nu} \phi_{\nu \rho} \phi^{\rho \s} \phi_{\s} \,, \quad
L^\3_8 = \phi_{\mu}  \phi^{\mu \nu} \phi_{\nu \rho} \phi^{\rho}\, \phi_{\s} \phi^{\s \lambda} \phi_{\lambda} \,,  \nb \\
&& L^\3_9 = \Box \phi \left(\phi_{\mu} \phi^{\mu \nu} \phi_{\nu}\right)^2  \,, \quad
L^\3_{10} = \left(\phi_{\mu} \phi^{\mu \nu} \phi_{\nu}\right)^3 \,.
\eea

Introducing the auxiliary variable $A_\mu$ as in (\ref{lagmult}), the general action \eqref{general action} becomes
\begin{eqnarray}\label{newform}
S[g,\phi;A_\mu,\lambda^\mu]   =  \int d^4 x \, \sqrt{-g} &&  \!\!\!\!\!\! \left( f_2\, R + C_\2^{\mu\nu\rho\sigma} \nabla_\mu A_\nu \, \nabla_\rho A_\sigma \right. \label{newform} \\
&&  \!\!\!\!\!\! \left. + f_3 \, {G}_{\mu\nu} \nabla^\mu A^\nu + 
C_\3^{\mu\nu\rho\sigma\alpha\beta}  \nabla_\mu A_\nu \nabla_\rho A_\sigma \nabla_\alpha A_\beta + 
\lambda^\mu (\phi_\mu - A_\mu)\right)\,, \nb
\end{eqnarray}
where the tensors $C_\2^{\mu\nu\rho\sigma}$ and $C_\3^{\mu\nu\rho\sigma\alpha\beta}$  are now expressed in 
terms of $A_\mu$ and $\phi$. Clearly, the two Lagrangians (\ref{general action}) and (\ref{newform}) are  equivalent.
 
\smallskip

Although we do not perform explicitly a Hamiltonian analysis here\footnote{All the details about the complete Hamiltonian analysis of quadratic theories can be found in  \cite{LN2}.}, let us briefly  comment  about  the role of the Lagrangian multipliers $\lambda^\mu$ and the relations they enforce.
Since the action~(\ref{newform}) does not involve the velocities of $\lambda^\mu$,  the corresponding conjugate momenta $p_\mu$  appear in the total Hamiltonian $H_T$ as primary constraints that weakly vanish.
The evolution of $p_i$ gives the secondary constraints $\phi_i - A_i \approx 0$. By contrast, the evolution of $p_0$ allows one to solve for the multiplier used in $H_T$ to impose the other primary constraint $\pi - \lambda^0\approx 0$, where $\pi$ is the momentum of~$\phi$.
The evolution of $\pi - \lambda^0$, on the other hand, fixes the multiplier associated with $p_0$.
All these constraints are {\it second class} and therefore can be consistently imposed in the Hamiltonian analysis; in particular the constraints $\phi_i - A_i \approx 0$ enables us to eliminate the velocity of $A_i$ in favour of the spatial derivative of $A_0$, as explained in detail in the next section.
It is thus clear that the constraints that follow from the $\lambda^\mu$ in (\ref{newform}) do not get mixed up with the (potential) extra primary constraint necessary to eliminate the \Ost mode,  which characterises degenerate theories.

\subsection{Covariant ADM decomposition}
\label{secADM}
\no 
In order to study the kinetic structure of the action (\ref{newform}), we must perform a $3+1$ decomposition of its building blocks. 
We now assume the existence of an arbitrary  slicing of spacetime with 3-dimensional spacelike hypersurfaces.   
We introduce the unit vector $n^\mu$ normal to the spacelike hypersurfaces, 
which is time-like and satisfies the normalization condition $n_\mu n^\mu=-1$.  This induces a three-dimensional metric, corresponding to the projection tensor on the spatial hypersurfaces,  defined by
\beq
h_{\mu\nu}\equiv g_{\mu\nu}+n_\mu n_\nu \,.
\eeq
Following the construction of \cite{LN1}, we 
define the spatial and normal projection of $A_\mu$, respectively
 \beq
 \tA_\mu\equiv h_\mu^\nu A_\nu \, ,
\qquad
 \An\equiv A_\mu n^\mu\,.
 \eeq
Let us now introduce the time direction vector $t^\mu=\partial/\partial t$ associated with a time coordinate $t$ that labels the slicing of spacelike hypersurfaces.
One can always decompose $t^\mu$ as 
\beq
t^\mu =N n^\mu +N^\mu,
\eeq
thus defining the lapse function $N$ and the shift vector $N^\mu$ orthogonal to $n^\mu$. We also define 
the ``time derivative'' of any  spatial tensor as the spatial projection of its Lie derivative  with respect to $t^\mu$. In particular, we have
\beq
\dotAn \equiv t^\mu \nabla_\mu\An \,, \qquad \dot\tA_\mu \equiv h_\mu^\nu{\cal L}_t \tA_\nu \,, \qquad
\dot h_{\mu\nu} \equiv h_\mu^\alpha h_\nu^\beta{\cal L}_t h_{\alpha\beta} \,.
\eeq
Due to the symmetric property of $\nabla_\mu A_\nu=\nabla_\nu A_\mu$, it is possible to express $\dot\tA_\mu$
in terms of $D_\mu A_*$ and $\dot h_{\mu\nu}$, therefore the only velocities (time derivative
of the fields) involved in $\nabla_\mu A_\nu$ are
\beq
\dotAn = N\, V_* + N^\mu D_\mu A_* \,,\qquad 
\dot h_{\mu\nu} =  2 \left( N K_{\mu\nu} + D_{(\mu}N_{\nu)} \right)\,, 
\eeq
where $V_* \equiv n^{\mu} \nabla_{\mu} A_*$\,, \,$K_{\mu\nu}$ is the extrinsic curvature tensor and $D_\mu$ denotes the 3-dimensional  covariant derivative associated with the spatial metric $h_{\mu\nu}$.

Instead of using the  velocities $\dot h_{\mu\nu}$ and $\dot A_*$, it is convenient to work with the covariant objects $K_{\mu\nu}$ and $V_*$ and interpret them as ``covariant velocities'' associated with the fields $h_{\mu\nu}$ and $A_*$. Working with these covariant quantities  allows us to avoid dealing with the lapse and the shift vector.

Using these  definitions, as well as the property $\nabla_\mu A_\nu=\nabla_\nu A_\mu$, the 3+1 covariant decomposition of  $\nabla_\mu A_\nu$ is given by 
\bea \label{decomposition_DA}
\nabla_{\mu} A_{ \nu} = 
D_{\mu} \hat A_{\nu}-A_*\, K_{\mu\nu}+ 2 \left( n_{(\mu} K_{\nu) \rho } \hat{A}^\rho- n_{(\mu}  D_{\nu)}A_*\right)
+\,n_\mu n_\nu \left(V_*- \hat{A}_\rho \,a^\rho \right) \,,
\eea
where $a^\mu\equiv n^\nu\nabla_\nu n^\mu$ is the acceleration vector.
One can rewrite  \eqref{decomposition_DA}  as
\beq
\label{decomposition_Aab}
\nabla_\mu A_\nu= \lambda_{\mu\nu}\, V_* +\Lambda_{\mu\nu}^{\ \ \rho\sigma} \, K_{\rho\sigma} + D_\mu\tA_\nu - 2\, n_{(\mu} D_{\nu)}\An - \lambda_{\mu\nu} \hat{A}_\rho \,a^\rho,
\eeq
with
\bea
\lambda_{\mu\nu}\equiv  n_\mu n_\nu\,, \qquad \Lambda_{\mu\nu}^{\ \ \rho\sigma}\equiv -\An \, h_{(\mu}^\rho h_{\nu)}^\sigma+2\, n_{(\mu} h_{\nu)}^{(\rho} \tA^{\sigma)} \, .
\eea
These two tensors fully characterise the velocity structure of the building block $\nabla_\mu A_\nu$ that appears in the action \eqref{newform} and will play an essential role in deriving the degeneracy conditions.

 \subsection{Horndeski Lagrangians and kinetic structure of the action}
 \no 
As an example of theories of the type (\ref{general action}),  and as a useful step for the general case, let us  first consider the particular case of the so-called quartic and quintic Horndeski Lagrangians\footnote{We use the subscripts $4$ and $5$, referring to `quartic' and `quintic', only for the Horndeski Lagrangians themselves. 
  According to our terminology,
 for all other associated variables  we use instead the subscripts $\2$ and $\3$ referring to `quadratic' or `cubic' types of  theory.}
 \bea
L_4^{\rm H} & = & f_2 R - 2 f_{2X} (\Box \phi^2 - \phi^{\mu\nu} \phi_{\mu\nu}) \,,
\label{quarticH}
\\
L_5^{\rm H} & = & f_3 G_{\mu\nu} \phi^{\mu\nu} + \frac{1}{3} f_{3X}(\Box \phi^3 - 3 \Box \phi \phi_{\mu\nu} \phi^{\mu\nu}
+ 2\phi_{\mu\nu} \phi^{\mu\sigma} \phi^\nu_{\, \sigma}) \,, \label{quinticH}
\eea
which correspond, respectively, to a quadratic and a cubic Lagrangian in our terminology.  Indeed, they are of the form (\ref{general action}), with 
\bea
a_1= - a_2 =  2 f_{2X} \,, \qquad a_3 = a_4 = a_5 =0 \,,
\eea
and 
\bea
3 \b_1= - \b_2 =\frac32 \b_3 = f_{3X} \,, \qquad  \b_i = 0 \,\,\,(i=4,...,10) \,.
\eea

It is instructive to extract the kinetic part of these two Lagrangians as the result will be useful for the general case.
The kinetic structure of the original Lagrangians \eqref{quarticH} and \eqref{quinticH} is the same as the following ones 
\bea
\label{H_kin}
L_{4\, \text{kin}}^{\rm H}  \, = \, C_{\2{\rm H}}^{\mu\nu\rho\sigma} \phi_{\mu\nu} \, \phi_{\rho\sigma}  \qquad \text{and}  \qquad 
L_{5\, \text{kin}}^{\rm H} \, = \,  C_{\3 {\rm H}}^{\mu\nu\rho\sigma\alpha\beta}  \phi_{\mu\nu} \, \phi_{\rho\sigma}  \,  \phi_{\alpha \beta} \,,
\eea
with\footnote{Note that $h^{\mu [\nu} h^{\rho]\sigma} $ denotes the anti-symmetrisation  on $(\nu,\rho)$
and $h^{\mu[\nu} h^{\vert \rho \vert \sigma} h^{\vert \alpha \vert\beta]}$ denotes the anti-symmetrisation on the second index of each
tensor $h$. The same notation holds for the terms defined with  the projector $\hat{P}$.}
\bea
 C_{\2{\rm H}}^{\mu\nu\rho\sigma} & = & \frac{2}{A_*^2}\left[ - f_2 \, h^{\nu [\mu} h^{\rho]\sigma} + 2 f_{2X} \left( A^2  h^{\nu[\mu} h^{\rho]\sigma} - \hat{A}^2 \hat{P}^{\nu[\mu} \hat{P}^{\rho]\sigma} \right) \right]  \, , \label{C2H} \\
C_{\3{\rm H}}^{\mu\nu\rho\sigma\alpha\beta} & = & - \frac{2 f_{3X}}{A_*^2} \left( A^2 h^{\mu[\nu} h^{\vert \rho \vert \sigma} h^{\vert\alpha \vert\beta]} - \hat{A}^2 \hat{P}^{\mu[\nu} \hat{P}^{\vert \rho \vert\sigma} \hat{P}^{\vert \alpha \vert \beta]} \right)  \, , \label{C3H}
\eea
where $A^2\equiv A_\mu A^\mu$, $\hat A^2\equiv  \hat A_\mu \hat A^\mu$ and we have introduced 
the projection tensor (orthogonal to the directions $n^\mu$ and $\hat{A}^\mu$)
\begin{equation}
\hat{P}_{\mu\nu} \equiv h_{\mu\nu} - \frac{1}{\hat{A}^2} \hat{A}_{\mu} \hat{A}_{\nu} \,.
\label{Phat}
\end{equation}
Notice that the tensors (\ref{C2H}) and (\ref{C3H}) are orthogonal to the vector $n^\mu$, therefore the kinetic terms do not contain the velocity $V_*$. 
This is the peculiarity of Horndeski Lagrangians which reflects in second order equations of motions.
\smallskip

Let us now turn to the general action \eqref{general action}.  In order to extract its kinetic part, 
it is convenient to re-express the curvature terms in the action as Horndeski Lagrangians so that 
one can use  the results above. The action \eqref{general action}  is thus rewritten as 
\bea\label{action with Horndeski}
S[g,\phi] = \int d^4x \, \sqrt{ -g } \, \left( L_4^{\rm H}  +
 \tC_\2^{\mu\nu\rho\sigma} \,  \phi_{\mu\nu} \, \phi_{\rho\sigma}
 +  L_5^{\rm H} + 
\tC_\3^{\mu\nu\rho\sigma\alpha\beta} \,  \phi_{\mu\nu} \, \phi_{\rho\sigma}  \,  \phi_{\alpha \beta}
\right) \,,
\eea
where the tensors $\tC_\2^{\mu\nu\rho\sigma} $ and $\tC_\3^{\mu\nu\rho\sigma\alpha\beta} $ are of the form \eqref{C2}-\eqref{C3} with the new functions 
\bea
&& \tilde a_1 = \aa - 2 f_{2X} \, , \qquad  \tilde a_2 = \ab + 2 f_{2X} \, ,  \\
&& \tilde\b_1 = \b_1 - \frac{1}{3} f_{3X} \, ,\qquad \tilde \b_2 = \b_2 + f_{3X} \, ,\qquad \tilde\b_3 =
\b_3 - \frac{2}{3} f_{3X} \,,
\eea
while  all the other functions remain unchanged. 

Replacing the Lagrangians $L_4^{\rm H} $ and $ L_5^{\rm H}$ in \eqref{action with Horndeski} with the kinetically equivalent ones (\ref{H_kin}), one finds that the kinetic structure of the total action is described by the tensors
\bea
 C_{\2}^{\mu\nu\rho\sigma}  =  C_{\2{\rm H}}^{\mu\nu\rho\sigma} + \tC_\2^{\mu\nu\rho\sigma}\,  \quad 
 \text{and}
 \quad
C_{\3}^{\mu\nu\rho\sigma\alpha\beta}  =C_{\3 {\rm H}}^{\mu\nu\rho\sigma\alpha\beta} +  \tC_\3^{\mu\nu\rho\sigma\alpha\beta} \, .
\eea
Only these tensors are relevant for the degeneracy conditions, which we derive below.

\subsection{Degeneracy conditions and primary constraints}
\no 
We now introduce the Hessian matrix of the Lagrangian with respect to the velocities $V_*$ and $K_{ij}$. 
This matrix  can be written in the form (introducing a factor $1/2$ for convenience)
 \bea
\mathbb{H} =
\left(
\begin{array}{cc}
\A & \B^{ij} \\
\B^{kl} & {\K}^{ij,kl}
\end{array}
\right) \,,
\eea
with
\beq
\A\equiv\frac12\, \frac{\partial^2 L}{\partial V_*^2} \,,\qquad
\B^{ij}\equiv\frac12\, \frac{\partial^2 L}{\partial V_* \partial K_{ij}}
\,,\qquad
\K^{ij,kl}\equiv\frac12\, \frac{\partial^2 L}{\partial K_{ij}\partial K_{kl}} \, . \label{ABK}
\eeq
The degeneracy of the theory is associated with the degeneracy
of its Hessian matrix, i.e. $\text{det} \mathbb{H} =0$. Equivalently, one can find a non trivial null eigenvector $(v_0, \V_{kl})$ such that 
\beq
\label{degenerate}
v_0\A+\B^{kl}\V_{kl}=0\,, \qquad v_0 \B^{ij}+\K^{ij,kl} \V_{kl}=0\,.
\eeq
These conditions translate into the existence of a primary constraint, which takes the form
\beq
v_0 \, \pi_* + \V_{ij} \pi^{ij} + \dots \approx 0 \,, \label{primaryconst}
\eeq
where we have introduced the ``covariant momenta'' conjugated respectively to $A_*$ and $h_{\mu\nu}$,
\beq
\pi_* \equiv  \frac{\delta {L}}{\delta V_*} \,, \qquad
\pi^{ij} \equiv \frac{\delta {L}}{\delta K_{ij}} \,,
\eeq
and 
the dots indicate  momentum-independent terms, involving only the  fields and their spatial derivatives.
Note that  we will always assume $v_0\neq 0$ since we are interested in removing the Ostrogradsky mode: therefore in the following we will fix $v_0=1$ without loss of generality.

It is important to keep in mind that the primary constraint (\ref{primaryconst}) is a {\it scalar} constraint involving only the scalar components of $\pi^{ij}$, i.e. $\V_{ij}\pi^{ij}$. It is indeed responsible for removing the scalar Ostrogradsky mode.
However, there could still be extra primary constraints in the vector sector of $\pi^{ij}$, which  can further reduce the number of degrees of freedom (dof) 
(as  pointed out in \cite{Achour:2016rkg}
and further stressed in \cite{deRham:2016wji}).
Indeed, as we will show  in what follows, some classes of theories that possess  the constraint (\ref{primaryconst}), also enjoy the two following primary constraints:
\beq
\tA_i \hat{P}_{jk} \pi^{ij} + \dots \approx 0 \,, \label{primaryvec}
\eeq
where we have used 
the projector (\ref{Phat}).
These constraints remove the two helicity-2 dof present in the metric sector, leaving the theory with only one dof.

\medskip
In order to compute the Hessian matrix of \eqref{action with Horndeski}, one needs to keep all terms quadratic and cubic in the velocities.
The Hessian matrix decomposes into its quadratic and  cubic contributions denoted 
\beq
\mathbb{H}_\2 =
\left(
\begin{array}{cc}
\A_\2 & \B_\2^{ij} \\
\B_\2^{kl} & \K_\2^{ij,kl}
\end{array}
\right) \, , \qquad 
\mathbb{H}_\3 =
\left(
\begin{array}{cc}
\A_\3 & \B_\3^{ij} \\
\B_\3^{kl} & \K_\3^{ij,kl}
\end{array}
\right) \,,
\eeq
with
\bea
 \A_\2 \equiv C_{\2}^{\mu\nu\rho\sigma}\,\lambda_{\mu\nu} \,\lambda_{\rho\sigma} \, , \qquad 
\B_\2^{ij}\equiv C_{\2}^{\mu\nu\rho\sigma} \,\lambda_{\mu\nu} \,\Lambda_{\rho\sigma}^{ ij} \, , \qquad 
\K_\2^{ij,kl}\equiv  C^{\mu\nu\rho\sigma}_{\2} \,\Lambda_{\mu\nu}^{ij} \,\Lambda_{\rho\sigma}^{ kl} \, , 
\label{ABK2}
\eea
\bea
\A_\3 \equiv 3\, C^{\mu\nu\rho\sigma\alpha\beta}_{\3}\,\lambda_{\mu\nu} \,\lambda_{\rho\sigma}\, \phi_{\alpha\beta} 
\, , \quad 
\B_\3^{ij}\equiv 3\, C^{\mu\nu\rho\sigma\alpha\beta}_{\3} \,\lambda_{\mu\nu} \,\Lambda_{\rho\sigma}^{ij} \,\phi_{\alpha\beta} \, , \quad
 \K_\3^{ij,kl}\equiv  3\, C^{\mu\nu\rho\sigma\alpha\beta}_{\3} \,\Lambda_{\mu\nu}^{ ij} \,\Lambda_{\rho\sigma}^{ kl} \, \phi_{\alpha\beta} \, . \nb \\
 \label{ABK3}
\eea

Introducing the tensor 
\bea\label{L}
L_{\mu\nu} & \equiv & \lambda_{\mu\nu} + \Lambda_{\mu\nu}^{ij} \V_{ij}  \,,
\eea 
the conditions \eqref{degenerate}  (with  $v_0=1$) for purely quadratic theories read 
\begin{eqnarray}\label{degeneracy_C2_intermsofC}
C_{\2}^{\mu\nu\rho\sigma} \lambda_{\mu\nu}  L_{\rho\sigma}  \; = \; 0 \,,  \qquad 
C_\2^{\mu\nu\rho\sigma} \Lambda_{\mu\nu}^{ij} L_{\rho\sigma} \; =  \; 0 \, .
\end{eqnarray}

On the other hand, the cubic Hessian matrix contains velocities. Therefore, the degeneracy conditions must be satisfied for arbitrary values of $\phi_{\alpha\beta}$. This implies that $\phi_{\alpha\beta}$ can  be ``factorised'' and the conditions \eqref{degenerate} in the cubic case are analogous  to the quadratic ones, namely
\begin{eqnarray}
\label{degeneracy_C3_intermsofC}
C_\3^{\mu\nu\rho\sigma\alpha\beta} \lambda_{\mu\nu}  L_{\rho\sigma}  \; = \; 0 \; , \qquad
C_\3^{\mu\nu\rho\sigma\alpha\beta} \Lambda_{\mu\nu}^{ij} L_{\rho\sigma} \; =  \; 0 \, .
\end{eqnarray}
The above equations mean that, in order to get a degenerate Lagrangian,   the projections of the tensors $C_{\2}^{\mu\nu\rho\sigma} L_{\rho\sigma}$ or $C_{\3}^{\mu\nu\rho\sigma\alpha\beta} L_{\rho\sigma}$, respectively via $\lambda_{\mu\nu}$ and $\Lambda_{\mu\nu}^{ij}$,   must vanish. 
As shown in Appendix~\ref{Tensorialstructure}, this implies that these tensors are necessarily of the form 
\bea
C_{\2}^{\mu\nu\rho\sigma} L_{\rho\sigma}  = 2 \m A^{(\mu} \hat{A}^{\nu)}   \,, \label{C2L}
\eea
and 
\bea
C_\3^{\mu\nu\rho\sigma\alpha\beta} L_{\alpha\beta} = 4 \m_1 \,  A^{(\mu} h^{\nu)(\rho} A^{\sigma)} + 4\m_2 \,  A^{(\mu}\tA^{\nu)} A^{(\rho}\tA^{\sigma)}   \,, \label{C3L}
\eea
where $\m$, $\m_1$ and $\m_2$ are arbitrary scalar quantities. 

By solving the conditions (\ref{C2L}), one recovers the quadratic theories identified in \cite{LN1}; they are summarised in Appendix \ref{app_quadratic}. Conditions (\ref{C3L}) are  solved in detail in Appendix \ref{app_cubic} and in the next section we report the various classes of purely cubic theories. Then, we will consider the possibility to merge  quadratic and cubic theories. In this case the additional condition to impose is that $L_{\mu\nu}$ is  the same in (\ref{C2L}) and (\ref{C3L}), i.e. we have to use the same $\V_{ij}$.

\section{Classification of cubic theories}
\no 
The degeneracy conditions for quadratic theories (i.e. with $f_3=b_i=0$) have already been solved and the corresponding theories identified in \cite{LN1}.  These quadratic theories were then examined in more details in \cite{Crisostomi:2016czh,Achour:2016rkg,deRham:2016wji}. 
In this section we thus focus our attention on the purely cubic theories, i.e. characterized by $f_2=a_i=0$. Solving the degeneracy conditions here is much more involved than in the quadratic case and rewriting them in the tensorial form \eqref{degeneracy_C3_intermsofC} is instrumental to obtain the full classification. Below, we simply present the full classification, indicating for each class the free functions among the $b_i$ and the constraints satisfied by the other functions.
All  the cubic theories we identify are summarised at the end of the section in Table~\ref{classes}. 
The details of how we have identified these classes are given in Appendix \ref{app_cubic}, where the reader can also find the explicit expression of the null eigenvectors associated with the degeneracy. The latter are indispensable to identify the healthy combinations of quadratic and cubic Lagrangians, which will be given in the next section. 

\subsection{Minimally coupled theories}
\no We start with the minimally coupled case, corresponding to $f_3=0$.
There are seven different classes of theories.

\renewcommand{\theenumi}{\alph{enumi}}

\hspace{1cm}\\
$\blacktriangleright$ {\bf $^3$M-I}: 
Four free functions $\b_1,\b_2,\b_3$ and $\b_4$ (with  $9\b_1+ 2 \b_2 \neq 0$). All the other functions are determined as follows:
\bea
\b_5 &=& - \frac{2}{X} \b_2 \, , \qquad \b_6  =  \frac{9 \b_1 \b_3 + 3 \b_4 X(\b_2 + \b_3) - 2 \b_2^2}
{X (9\b_1 + 2 \b_2)} \,,\nb \\
\b_7 &= &- \frac{3}{X} \b_3 \,, \qquad 
\b_8  =  \frac{9 \b_1 \b_3 - 6 \b_4 X(\b_2 + \b_3) + 6 \b_2 \b_3 + 4 \b_2^2}
{X^2 (9 \b_1 + 2 \b_2)} \,,\nb\\
\b_9 & = & \frac{1}{X^2 (9 \b_1 + 2 \b_2)^2} \Big[3 \b_4^2 X^2 (9 \b_1 + 3 \b_2 + \b_3) - 2 \b_4 X \left(9 \b_1 (\b_2 - \b_3) + 4 \b_2^2\right) \nb \\
&& + 24 \b_1 \b_2^2 + 54 \b_1^2 \b_2 + 27 \b_1^2 \b_3 + 4 \b_2^3\Big] \,,\nb \\
\b_{10} & = & \frac{1}{X^3 (9 \b_1 + 2\b_2)^3} \Big[3 \b_4^3 X^3 (9 \b_1 + 3 \b_2 + \b_3) - 6 \b_2 \b_4^2 X^2(9 \b_1 + 3 \b_2 + \b_3) \nb \\
&& + 2 \b_4 X \left(81 \b_1^2 (\b_2 + \b_3) + 18 \b_1 \b_2 (3 \b_2 + 2 \b_3) + 2 \b_2^2 (5 \b_2 + 3 \b_3)\right) \nb \\
&& - 2\left(54 \b_1^2 \b_2 (\b_2 + 2 \b_3) + 4 \b_1 \b_2^2 (7 \b_2 + 9 \b_3) + 81 \b_1^3 \b_3 + 4 \b_2^3(\b_2 + \b_3)\right) \Big] \,.\label{3M-I}
\eea

This class includes the pure quintic beyond Horndeski Lagrangian: 
\bea
L^{\rm bH}_{5} &=& 
f(\phi, \,X) 
\Big[
X \left( (\Box\phi)^3- 3\, \Box \phi\, \phi_{\mu \nu} \phi^{\mu \nu} 
+ 2 \phi_{\mu \nu}\phi^{\nu \rho} \phi^{\mu}_{\rho}
\right) \label{quinticBH} \\
&-& 3 \left( 
\left(\Box \phi\right)^2 \phi_{\mu} \phi^{\mu \nu} \phi_{\nu} 
- 2\, \Box \phi\, \phi_{\mu}  \phi^{\mu \nu} \phi_{\nu \rho}
\phi^{\rho} - \phi_{\mu \nu} \phi^{\mu \nu} \phi_{\rho} \phi^{\rho \s}
\phi_{\s} + 2 \phi_{\mu} \phi^{\mu \nu} \phi_{\nu \rho} \phi^{\rho \s} \phi_{\s}  \right) \Big] \,, \nb
\eea
which corresponds to the choice of functions
\begin{equation}
\frac{b_1}{X} = - \frac{b_2}{3X}  = \frac{b_3}{2X} = - \frac{b_4}{3} = \frac{b_5}{6} = \frac{b_6}{3} = - \frac{b_7}{6} = f \,. 
\end{equation}
The above combination is special as it leaves the Lagrangian linear in $V_*$, therefore in (\ref{ABK3})  $\A_3 = 0$.

Notice that in this class  $9\b_1+ 2 \b_2 \neq 0$. The condition $9\b_1+ 2 \b_2 =0$ leads to the next three classes.

\hspace{1cm}\\
$\blacktriangleright$ {\bf $^3$M-II}:
Three free functions $\b_1,\b_3,\b_6$ (with $9 \b_1 - 2 \b_3 \neq 0$). All the other functions are given by
\bea
\b_2&=&-\frac{9}{2} \b_1 \, , \qquad
\b_4 = -\frac{3}{X} \b_1 \, , \qquad
\b_5 = \frac{9}{X} \b_1 \, , \nb \\
\b_7 &=& - \frac{3}{X} \b_3 \,, \qquad
\b_8=\frac{3\b_3-2\b_6 X}{X^2} \, , \nb \\
\b_9 &=& \frac{9 \b_1 (\b_3 + 2 \b_6 X) - 81 \b_1^2 - 2 \b_6^2 X^2}{3 X^2 (9 \b_1 - 2 \b_3)} \, , \nb \\
\b_{10} &=& \Big[18 \b_6 X \left(-12 \b_1 \b_3 + 27 \b_1^2 + 2 \b_3^2\right) - 36 \b_3\left(-8 \b_1 \b_3 + 18 \b_1^2 + \b_3^2\right) \nb \\ 
&& - 12 \b_3 \b_6^2 X^2 + 4 \b_6^3 X^3\Big]
\Big[9 X^3 (9 \b_1 - 2 \b_3)^2\Big]^{-1} \,, \nb
\eea
In this class $9 \b_1 - 2 \b_3 \neq 0$. The case $9 \b_1 - 2 \b_3 = 0$ (and $9\b_1 + 2\b_2 = 0$) is described by the next two classes. 

\hspace{1cm}\\
$\blacktriangleright$ {\bf $^3$M-III}:
A single free function $\b_1$. All the other functions are determined in terms of $\b_1$ as follows:
\bea
\b_2 &=&-\frac92 \b_1 \, , \qquad \b_3 = \frac92 \b_1 \,, \qquad \b_4 = - \frac{3 \b_1}{X} \,, \qquad
\b_5 = \frac{9}{X} \b_1 \, , \qquad \b_6 = \frac{9 \b_1}{2X} \,, \nb \\
\b_7 &=& - \frac{27}{2X} \b_1 \,, \qquad \b_8  =  \frac{9 \b_1}{2X^2} \,, \qquad \b_9 = - \frac{3\, \b_1}{2X^2} \, , \qquad \b_{10} = - \frac{\b_1}{X^3} \, . \nb
\eea

\hspace{1cm}\\
$\blacktriangleright$  {\bf $^3$M-IV}:
Five free functions $\b_1,\b_4,\b_5,\b_8,\b_{10}$. The other functions are given by
\bea
&& \b_2=-\frac92 \b_1 \,, \qquad \b_3=\frac92 \b_1 \,, \qquad  \b_6=-3\b_4-\frac{9}{2X}\b_1 \,, \nb \\[2ex]
&&\b_7=-3\b_5 + \frac{27}{2X}\b_1 \,, \qquad \b_9= \frac{3 \b_1 - 2 X (2 \b_4 + \b_5)}{2 X^2} \,. \nb
\eea

\hspace{1cm}\\
$\blacktriangleright$ 
{\bf $^3$M-V}:
Two free functions, $\b_1$ and $\b_{4}$, while the other functions are given by
\bea
\b_2=\b_3=\b_5=\b_6=\b_7=\b_8=0 \,,\qquad 
\b_9 = \frac{\b_4^2}{3\b_1} \,,  \qquad \, \b_{10}=\frac{\b_4^3}{27 \b_1^2} \, .
\eea
There is only one (scalar) dof that propagates due to the primary constraints (\ref{primaryvec}),
and their associated secondary constraints.

\hspace{1cm}\\
$\blacktriangleright$  {\bf $^3$M-VI}:
Six free functions $\b_1$, $\b_4$, $\b_5$, $\b_8$, $\b_9$ and $\b_{10}$.
All the other functions vanish:
\bea
\b_2=\b_3=\b_6=\b_7=0 \,.
\eea
Again, there is only one (scalar) dof that propagates. 

\hspace{1cm}\\
$\blacktriangleright$  {\bf $^3$M-VII}:
Four free functions $\b_5$, $\b_7$, $\b_8$ and $\b_{10}$. The remaining functions vanish, except $\b_9$:
 \bea
\b_1=\b_2=\b_3=\b_4=\b_6=0\,, \qquad \b_9=- \frac{\b_5}{X}\,.
\eea

\subsection{Non-minimally coupled theories}
\no We now consider the purely cubic Lagrangians with $f_3 \neq 0$. There are two classes of theories.

\hspace{1cm}\\
$\blacktriangleright$ {\bf $^3$N-I:} 
In addition to $f_3$, the functions $\b_1$ and $\b_4$ are free (with the only restriction $\b_1\neq 0$). The other functions are determined as follows:
\bea
&& \b_2=  -3\,\b_1 \, , \qquad \b_3 = 2\,\b_1 \, , \qquad \b_6 = -\b_4 \, , \nb \\[2ex]
&& \b_5 = \frac{2 (f_{3X} - 3 \b_1)^2 - 2 \b_4 f_{3X} X}{3 \b_1 X} \, ,\qquad \b_7 = \frac{2 \b_4 f_{3X} X - 2 (f_{3X} - 3 \b_1)^2}{3 \b_1 X} \, , \nb \\
&& \b_8 = \frac{2 (3 \b_1 + \b_4 X - f_{3X})\left((f_{3X} - 3 \b_1)^2 - \b_4 f_{3X} X\right)}
{9 \b_1^2 X^2} \,, \nb \\
&& \b_9 = \frac{2 \b_4 (3 \b_1 + \b_4 X - f_{3X})}
{3 \b_1 X} \, ,\qquad \b_{10} = \frac{2 \b_4 (3 \b_1 + \b_4 X - f_{3X})^2}
{9 \b_1^2 X^2} \, . \label{3N-I}
\eea
Quintic Horndeski (\ref{quinticH}), as well as the combination of quintic Horndeski plus quintic beyond Horndeski (\ref{quinticBH}), is included in this class of models.

\hspace{1cm}\\
$\blacktriangleright$  {\bf $^3$N-II}:
Free functions $\b_5$, $\b_8$ and $\b_{10}$, in addition to $f_3$. The other functions are given by 
\bea\label{NMclass2}
&& \b_1 = \b_2= \b_3 =0 \,, \qquad  \b_7 = - \b_5 \,, \nb \\[2ex]
&& \b_4 = -\b_6= \frac{f_{3X}}{X} \, , \qquad \b_9 = -\frac{2\,f_{3X} + X \b_5 }{X^2} \, . \nb
\eea

\subsection{Minkowski limit}

\no Here we discuss which ones among the classes of theories described above
 admit a healthy Minkowski limit, i.e. the limit  where the metric is given by $g_{\mu \nu} = \eta_{\mu \nu}$ and the metric fluctuations are ignored.
In this limit, 
only the scalar sector is dynamical and the Hessian matrix reduces to its purely scalar component, i.e. $\A$. For cubic theories, the degeneracy is thus expressed by the condition $\A_\3 =0$, which imposes the relations
\begin{equation}
b_1= - \frac{b_2}{3} = \frac{b_3}{2} \,, \qquad b_4 = - \frac{b_5}{2} = -b_6 = \frac{b_7}{2} \,, \qquad b_8=b_9=b_{10}=0 \,. 
\label{flatcon}
\end{equation}
The only classes that  satisfy these conditions  are
\begin{itemize}
	\item $^3$M-I:  Beyond Horndeski theory,
	\item $^3$N-I: Beyond Horndeski and Horndeski theory, 
	\item $^3$N-II: Imposing also $b_5= - 2 f_{3X}/X$ and $b_8=b_{10}=0$ .
\end{itemize}
This shows that there is a new theory, $^3$N-II, which  propagates three degrees of freedom on curved spacetime and has a healthy Minkowski limit.
On the other hand, theories that do not satisfy (\ref{flatcon}) could still have a healthy decoupling limit 
around a non-trivial background.

\bigskip

 \begin{table}[H]
 	\centering
 	\begin{tabular}{ | c || c | c | c | c | }
 		\hline
 		\multicolumn{5}{|c|}{Minimally coupled theories} \\
 		\hline
 		Classification & \quad $\#$ dof \, $$  & Free functions & Minkowski limit &Examples\\
 		\hline
 		$^3$M-I   & 3 &  $i$=1,2,3,4 & \checkmark\, (bH)  & bH, $\Omega\otimes$bH${}^{(1)}$  \\
 		$^3$M-II  & 3 & $i$=1,3, 6 & X & \\
 		$^3$M-III & 3 & $i$=1 & X  &  \\ 
 		$^3$M-IV  & 3 & $i$=1,4,5,8,10 & X & \\
 		$^3$M-V  & 1 & $i$=1,4 & X & \\
 		$^3$M-VI  & 1 & $i$=1,4,5,8,9,10& X &  \\
 		$^3$M-VII  & 3 & $i$=5,7,8,10 & X & \\
 		\hline \hline
 		\multicolumn{5}{|c|}{Non-minimally coupled theories} \\
 		\hline
 		Classification & \quad $\#$ dof \, $$ & Free functions & Minkowski limit &Examples\\
 		\hline
 		$^3$N-I & 3 & $f_3$, $i$=1,4  & \checkmark\, (H, bH) & H, H+bH, $\Gamma\otimes$H${}^{(2)}$, $(\Omega,\Gamma)\otimes$H${}^{(3)}$ \\
 		$^3$N-II  & 3 & $f_3$, $i$=5,8,10  & \checkmark  &  \\ 
 		\hline
 	\end{tabular}
 	\caption{Summary of all cubic degenerate classes. The subscript $i$ in free functions indicates which functions among $b_i$ are free. Examples: 
    $(1)$: Theories obtained by the generalised conformal transformation ($\Omega$) from beyond Horndeski (bH). $(2)$: Theories obtained by the generalised disformal transformation ($\Gamma$) from Horndeski (H). This is equivalent to a combination of Horndeski and beyond Horndeski \cite{Crisostomi:2016tcp}. $(3)$: Theories obtained by the generalised conformal and disformal transformation from Horndeski. See section V for discussions about the generalised conformal and disformal transformation.     
    }\label{classes}
 \end{table}

\section{Merging quadratic with cubic theories}
\no In this section we wish to determine all the theories of the form (\ref{general action}), i.e. quadratic plus cubic Lagrangians, that are degenerate.
Adding two degenerate Lagrangians does not always yield a degenerate one. This is the case only if the null eigenvectors associated with the two Lagrangians coincide. Therefore, in order to see whether the combination of two Lagrangians is viable, one  needs to compare their eigenvectors, which are all listed in Appendix  \ref{app_quadratic} for quadratic theories and in Appendix \ref{app_cubic} for cubic ones, and check when they are equal. 

We present four tables describing all the different possibilities for merging quadratic and cubic
theories. 
We indicate with \checkmark  theories that can be freely combined,
with X theories that cannot be combined, and with $(n)$ theories that can be combined imposing the additional condition(s) $(n)$ listed below each table.

\hspace{1cm}\\
{\bf Minimally coupled quadratic plus minimally coupled cubic theories}

\begin{table}[H]
\centering
\begin{tabular}{ | c || c | c | c | c | c | c | c |  }
 		\hline
 		 & \quad $^3$M-I \, $$  & \quad $^3$M-II \, $$ & \quad $^3$M-III \, $$ & \quad $^3$M-IV \, $$ & \quad$^3$M-V \, $$ & \quad $^3$M-VI \, $$ & \quad $^3$M-VII \, $$ \\
 		\hline \hline
 		\, $^2$M-I \, $$  &  $(1)$ & $(2)$ & \checkmark & X & $(3)$ & X & X \\
 		\,  $^2$M-II \, $$ &  X  & X & \checkmark & \checkmark & X & X & $(4)$ \\
 		\, $^2$M-III \, $$ & X & X & X & X & \checkmark & \checkmark & $(5)$ \\ 
 		\hline
 	\end{tabular}
 \label{tableI2M+3M}
 \end{table}

\no{{Conditions:}}
\renewcommand{\theenumi}{(\arabic{enumi})}

\begin{enumerate}

\item $\b_4 = \frac{-6 a_1 \b_1 + 4 a_2 \b_2 + a_3 X (9 \b_1 + 2 \b_2)}{2 X (a_1 + 3 a_2)}$

\item $\b_6 = \frac{3 (6 a_1 \b_1 + 4 a_2 \b_3 + a_3 X (2 \b_3 - 9 \b_1))}{4 X (a_1 + 3 a_2)}$ 

\item $\b_4 = -\frac{3 \b_1 ( 2 a_1 - 3 a_3 X)}{2 X (a_1  + 3 a_2)}$

\item $\b_7 = -3 \b_5 $

\item $\b_7 = 0$ (1 dof)

\end{enumerate}

Notice that condition (5) eliminates also the 2 tensor dof, leaving the joined classes $^2$M-III + $^3$M-VII with only one scalar dof.

The quartic beyond Horndeski theory $L_4^{\rm bH}$ is included in $^2$M-I, while the quintic beyond Horndeski theory $L_5^{\rm bH}$ (\ref{quinticBH}) is included $^3$M-I. They satisfy the condition (1) thus the combination $L_4^{\rm bH}$ + $L_5^{\rm bH}$ is still viable \cite{LN1,Crisostomi:2016tcp}.  

\hspace{1cm}\\
{\bf Non-minimally coupled quadratic plus minimally coupled cubic theories}
		\begin{table}[H]
	\centering
	\begin{tabular}{ | c || c | c | c | c | c | c | c |  }
		\hline
 		 & \quad $^3$M-I \, $$  & \quad $^3$M-II \, $$ & \quad $^3$M-III \, $$ & \quad $^3$M-IV \, $$ & \quad$^3$M-V \, $$ & \quad $^3$M-VI \, $$ & \quad $^3$M-VII \, $$ \\
 		\hline \hline
 		\, $^2$N-I \, $$ & (1) \& (3) & (1) \& (6) & (1) & X & (1) \& (4) & X & X \\
 		\,  $^2$N-II \, $$ & X & X & X & X & X & X & (7) \\
 		\, $^2$N-III \, $$ & (3) & (6) & \checkmark & X & (4) & X & X \\ 
		\, $^2$N-IV \, $$  & (2) \& (3) & (2) \& (6) & (2) & X & (5) & X & X \\
 		\hline
 	\end{tabular}
\label{tableI2N+3M}
 \end{table}

\no{{Conditions:}}
\begin{enumerate}

\item $a_3= -\frac{8 a_1 f_{2X}}{f_2}+\frac{6a_1 + 4 f_{2X}}{X}-\frac{4f_2}{X^2}$

\item $a_3= \frac{12 a_2 f_{2X}}{f_2}-\frac{8(a_2 -  f_{2X})}{X}-\frac{6f_2}{X^2}$

\item $\b_4= \frac{2 f_{2X} (9 \b_1 + 2 \b_2)}{f_2} - \frac{2 (6 \b_1 + \b_2)}{X}$

\item $\b_4= 6\b_1 \left(\frac{3 f_{2X}}{f_2} -\frac{2}{X}\right)$

\item $\b_4=\frac{3 \b_1 (X (a_3 X + 4 f_{2X}) - 2 f_2)}{2 X (a_2 X + f_2)}$

\item $\b_6= \frac{3 (6 \b_1 f_2 - 9 \b_1 f_{2X}X - \b_3 f_2 + 2 \b_3 f_{2X}X)}{f_2 X}$

\item $\b_7=-\b_5$

\end{enumerate}

The quartic Horndeski theory $L_4^{\rm H}$ (\ref{quarticH}) is included in $^2$N-I. The combination $L_4^{\rm H} + L_5^{\rm bH}$ does {\it not} satisfy the conditions (1) and (3), thus this combination is not degenerate \cite{LN1,Crisostomi:2016tcp}. 

\hspace{1cm}\\
{\bf Minimally coupled quadratic plus non-minimally coupled cubic theories}

\begin{table}[H]
\centering
\begin{tabular}{ | c || c  c  c | }
 		\hline
 		 & \quad $^2$M-I \, $$  & \quad $^2$M-II \, $$ & \quad $^2$M-III \, $$  \\
 		\hline \hline
 		\, $^3$N-I \, $$  & X & X & X \\
		\hline
 		\, $^3$N-II \, $$ & X & X & X \\
 		\hline
 	\end{tabular}
\label{tableI2M+3N}
 \end{table}

The quintic Horndeski theory $L_5^{\rm H}$ (\ref{quinticH}) is included in $^3$N-I. As can be seen from the table, it is not possible to combine $^3$N-I and $^2$M-I thus the combination $L_5^{\rm H} +L_4^{\rm bH}$ is not viable \cite{Crisostomi:2016tcp}. 

\newpage

\no {\bf Non-minimally coupled quadratic plus non-minimally coupled cubic theories}

\begin{table}[H]
\centering
\begin{tabular}{ | c || c c c c | }
 		\hline
 		 & \quad $^2$N-I \, $$ & \quad $^2$N-II \, $$ & \quad $^2$N-III \, $$ & \quad $^2$N-IV \, $$\\
 		\hline \hline
 		\, $^3$N-I \, $$  & $(1)$ & X & X & X \\
		\hline
 		\,  $^3$N-II \, $$ & X & \checkmark & X & X \\
 		\hline
 	\end{tabular}
 \label{tableI2N+3N}
 \end{table}

\no{{Conditions:}}
\begin{enumerate}
\item $\b_4 = \frac{-a_1 f_{3X} X - 6 \b_1 f_2 + 6 \b_1 f_{2X} X + 2 f_2 f_{3X}}{f_2 X}$

$a_3 =  \frac{2 \left(\b_1 \left(9 a_1 f_2 X -12 a_1 f_{2X} X^2 + 6 f_2 f_{2X} X - 6 f_2^2\right) + 2 f_{3X}(f_2 - a_1 X)^2\right)}{3 \b_1 f_2 X^2}$ 
\end{enumerate}

The classes $^2$N-I and $^3$N-I contain three free functions each, thus the combination $^2$N-I + $^3$N-I contains four free functions due to the conditions (1). In the next section, we show that this theory can be obtained by the generalised conformal and disformal transformation from $L_4^{\rm H}$ + $L_5^{\rm H}$.

\section{Conformal and disformal transformation}

\no  We now investigate which ones among  the cubic theories
can be obtained from known Lagrangians through conformal and disformal transformations. 
The same analysis for quadratic theories can be found in~\cite{Crisostomi:2016czh, Achour:2016rkg}.
First we identify the class of theories minimally coupled with gravity (i.e. $f_3=0$)  that can be obtained from beyond Horndeski (\ref{quinticBH}) by a conformal transformation.
Then,  we  study the class of theories that can be obtained from Horndeski theory (\ref{quinticH}) by a conformal together with a  disformal transformation.

\subsection{Conformal transformation on Beyond Horndeski} 

\no It was shown in \cite{Crisostomi:2016tcp} that under the generalised disformal transformation 
\begin{equation}
\bar{g}_{\mu \nu} = g_{\mu \nu} + \Gamma(X) \phi_{\mu} \phi_{\nu} \,,
\label{disconf}
\end{equation}
beyond Horndeski theory is transformed into itself:
\begin{equation}
\bar{L}_5^{\rm bH}[\bar{f}] = L_5^{\rm bH}[f], 
\end{equation}
where $f= \bar{f}/(1+ X \Gamma)^{7/2}$. On the other hand, under the generalised conformal transformation
\begin{equation}
\bar{g}_{\mu \nu} = \Omega(X) g_{\mu \nu} \,,
\end{equation}
it transforms as 
\begin{equation}
\bar{L}_5^{\rm bH}[\bar{f}] = L_5^{\rm bH}[f] + \sum_i \hat b_i L^\3_i \,,
\end{equation}
where 
\bea
&& f= \frac{\bar{f}}{\Omega^2}\,, \qquad \hat b_4= - \hat b_6 = \frac{3\,\bar{f}\, X \, \Omega_X}{\Omega^3}\,, \qquad \hat b_8= \frac{6\,\bar{f}\, \Omega_X}{\Omega^3}\,, \nb \\[2ex]
&& \hat b_9 =  \frac{6\,\bar{f}\, \Omega_X \left( X\, \Omega_X - \Omega \right)}{\Omega^4}\,, \qquad \hat b_{10} = \frac{6\,\bar{f}\, \Omega_X^2 \left( X\, \Omega_X - \Omega \right)}{\Omega^5}\,,
\eea
and the other $\hat b_i$ vanish. 
In terms of the total 
$b_i$, this gives 
\bea
&&b_1= X f\,, \quad b_2=-3X f\,, \quad b_3=2X f\,, \quad  b_4= - b_6 = -3\,f + \frac{3\,f\, X \, \Omega_X}{\Omega}\,, \quad b_5= - b_7 = 6\,f\,, \nb \\
&& b_8= \frac{6\,f\, \Omega_X}{\Omega}\,, \qquad b_9 =  \frac{6\,f\, \Omega_X \left( X\, \Omega_X - \Omega \right)}{\Omega^2}\,, \qquad b_{10} =  \frac{6\,f\, \Omega_X^2 \left( X\, \Omega_X - \Omega \right)}{\Omega^3}\,.
\label{BHconf}
\eea
These $b$'s satisfy conditions (\ref{3M-I}), thus this theory is included in class $^3$M-I.

\subsection{Conformal and disformal transformation on Horndeski}

\no The generalised conformal and disformal transformation 
\begin{equation}
\bar{g}_{\mu \nu} = \Omega(X) g_{\mu \nu} + \Gamma(X) \phi_{\mu} \phi_{\nu} \,, \label{confdistransf}
\end{equation}
transforms the Horndenski action as 
\begin{equation}
\bar{L}_5^{\rm H}[\bar{f}_3] = L_5^{\rm H}[f_3] + L_5^{\rm bH}[f] + \sum_i b_i L^\3_i \,,
\label{confdis}
\end{equation}
where
\bea
f_3 &=&   \frac{\bar{f}_3 \sqrt{\Om}}{ \sqrt{\Om + X \Ga}} + \int{\bar{f}_3\,\frac{\left(\Om - X \Om_X \right)\Ga + X \Om \Ga_X}{2\sqrt{\Om}\left( \Om + X \Ga \right)^{3/2}}\,d X} \,, \\[1ex]
f &=& \frac{\bar{f}_{3\bar{X}} \sqrt{\Omega } \left(\Omega_X + X \Gamma_X \right)}{3\left(\Omega + X \Gamma \right)^{5/2}} \,,\\[1ex]
b_4 &=& - b_6 = \frac{\bar{f}_{3\bar{X}} \Omega_X \sqrt{\Omega }}{\left(\Omega + X \Gamma \right)^{5/2}} \,, \\[1ex]
b_5 &=& -b_7 = \frac{2 \bar{f}_{3\bar{X}} \Omega_X \left[X \left(\Omega_X + X \Gamma_X \right) - \Omega \right]}{\sqrt{\Om}(\Omega + X \Gamma )^{5/2}} \,, \\[1ex]
b_8 &=& \frac{2 X \bar{f}_{3\bar{X}} \Omega_X \left[\Om_X \left(\Omega_X + X \Gamma_X \right) + \Omega \Ga_X \right]}{\Om^{3/2}(\Omega + X \Gamma )^{5/2}} \,, \\[1ex]
b_9 &=& - \frac{2 X \bar{f}_{3\bar{X}} \Omega_X \Ga_X }{\sqrt{\Omega }\left(\Omega + X \Gamma \right)^{5/2}} \,, \\[1ex]
b_{10} &=& - \frac{2 X \bar{f}_{3\bar{X}} \Omega_X^2 \Ga_X }{\Omega^{3/2}\left(\Omega + X \Gamma \right)^{5/2}} \,.
\eea
and the other $b_i$ vanish. One can check that this theory satisfies the conditions (\ref{3N-I}), thus it is included in class $^3$N-I. Theories in class $^3$N-I have three free functions. On the other hand, the action (\ref{confdis}) contains $\bar{f}_3$, $\Omega$ and $\Gamma$. Thus there is the same number of free functions. Indeed we can relate $\bar{f}_{3\bar{X}}$, $\Omega_X$ and $\Gamma_X$ to $f_{3X}$, $f$ and $b_4$ as 
\bea
\bar{f}_{3\bar{X}} &=& \frac{f_{3X} \left(\Omega + X \Gamma\right)^{5/2}}{\sqrt{\Omega}\left[\Omega - X \left(\Omega_X + X \Gamma_X \right)\right]} \,,\\[1ex]
\Omega_X &=& \frac{b_4 \Omega}{3 X f + f_{3X}} \,, \\[1ex]
\Gamma_X &=& \frac{\Omega \left(3 f - b_4\right) }{X \left(3 X f + f_{3X}\right)} \,.
\eea
Thus, theories in class $^3$N-I can be mapped to Horndeski if the transformation (\ref{confdistransf}) is invertible.

Finally we consider the generalised conformal and disformal transformation from $L_4^{\rm H}$ + $L_5^{\rm H}$. Using the result for the transformation of $L_4^{\rm H}$ obtained in \cite{Crisostomi:2016czh, Achour:2016rkg}, we can show that this theory corresponds to the combination of $^{2}$N-I and $^{3}$N-I and satisfies the condition (1). This theory has four free functions, which correspond to $\bar{f}_2, \bar{f}_3, \Omega$ and $\Gamma$. Thus this theory can be regarded as the ``Jordan frame" version of the Horndenski theory where the gravitational part of the Lagrangian is described by Hordenski with the metric $\bar{g}_{\mu \nu}$, $L_4^{\rm H}[\bar{g}]+L_5^{\rm H}[\bar{g}]$, while the matter is non-minimally coupled through $g_{\mu \nu}$. By performing the generalised conformal and disformal transformation, the gravitational action is described by the combination of $^{2}$N-I and $^{3}$N-I and the metric is minimally coupled to matter.  

\section{Conclusions}
\no In this paper, we presented all Ostrogradsky ghost-free theories that are at most cubic in the second derivative of the scalar field, and that propagate at most three degrees of freedom. Extending Horndeski's results, we have found new Lagrangians, which lead to higher order equations of motion but avoid Ostrogradsky instabilities by means of constraints that prevent the propagation of dangerous extra degrees of freedom.

In order to achieve our results, we used the degeneracy criterium introduced in \cite{LN1}, and classified the Lagrangians that are degenerate, i.e. whose Hessian matrix, obtained by taking the second derivatives of the Lagrangian with respect to velocities, is degenerate.
In total, we identified seven classes of minimally coupled cubic theories and two classes of non-minimally coupled cubic theories, which contain as subclasses all known scalar-tensor theories which are cubic in second derivatives of the scalar field. We also investigated which cubic theories admit a well-defined Minkowski limit, i.e. when the metric is frozen to its Minkowski value. Our results are summarised in the Table \ref{classes}.

We then studied in which cases it is possible to combine any of these cubic theories with the previously identified quadratic ones. 
Note that one can also add  arbitrary terms of the form $P(X,\phi)$ and $Q(X,\phi) \square \phi$ without changing the  degeneracy of the total Lagrangian. 
We confirmed the previous finding that the combination of 
quartic or quintic beyond Horndeski  with a different Horndeski is not viable. 
Finally, we studied whether our cubic theories are related to known Lagrangians through generalised conformal or disformal  transformations. We identified the theory, with four free functions, that is obtained by the generalised conformal and disformal transformation from
the combination of quartic and quintic Horndeski Lagrangians. 
   
\smallskip

Various interesting developments are left for the future. First, phenomenological aspects of these new theories should be investigated, in particular
studying the  existence of stable cosmological FLRW solutions -- possibly self-accelerating -- and their properties, by using for instance the effective description of dark energy (see e.g. \cite{Gleyzes:2014rba} for a review and \cite{Gleyzes:2015pma} for a recent generalization that includes non-minimal couplings to matter). It would also be worth analysing possible distinctive features of screening mechanisms in these set-ups. Secondly, on the theory side, it would be interesting to analyse further generalizations of scalar-tensor theories containing higher powers
of second derivatives of the scalar field. Such theories do not admit a well-defined Minkowski limit, and some explicit examples
have been discussed in \cite{Crisostomi:2016czh} and in \cite{Ezquiaga:2016nqo}. A more complete classification using the techniques we presented should be feasible, and left for future investigations.

\acknowledgements
KK is supported by the UK Science and Technologies Facilities Council grants ST/K00090X/1 and ST/N000668/1 and the European Research Council through grant 646702 (CosTesGrav). 
GT is partially supported by STFC grant ST/N001435/1.
MC and GT are grateful to CERN Theoretical Physics Department for hospitality and financial support during the development of this project.

\appendix

\section{Curvature dependent Lagrangians}
\label{Nterms}
\no Curvature tensors depend quadratically on the extrinsic curvature so,
according to the kinetic structure presented in Sec \ref{secADM}, their combination with the second derivative of the scalar field yields cubic powers in velocities. 
All the possible quadratic and cubic terms in velocities involving the curvature are
\beq
L_R=\sum_{i=1}^{2} L_2 [f_i] + \sum_{i=3}^{9} L_3 [f_i] \,,
\eeq
where
\bea
L_2 [f_1] = f_1\, R_{\mu\nu}\phi^\mu\phi^\nu \,, \qquad
L_2 [f_2] = f_2 \, R \,; \label{Nquad}
\eea
and
\bea
&& L_3 [f_3] = f_3 \, R_{\mu\nu} \phi^{\mu \nu} \,, \label{Ncub1} \\[1ex]
&& L_3 [f_4] = f_4 \, R \, \Box \phi  \,, \qquad\qquad\quad
L_3 [f_5] = f_5 \, R \,  \phi_{\mu} \phi^{\mu \nu} \phi_{\nu} \,,  \label{Ncub2} \\[1ex]
&& L_3 [f_6] = f_6 \, R_{\mu\nu} \phi^\mu \phi^\nu \, \Box \phi  \,, \qquad
L_3 [f_7] = f_7 \, R_{\mu\nu} \phi^\mu \phi^\nu \, \phi_{\rho} \phi^{\rho \sigma} \phi_{\sigma} \,,  \label{Ncub3} \\[1ex]
&& L_3 [f_8] = f_8 \, R_{\mu\nu} \phi^{\mu \rho} \phi_\rho \phi_\nu \,, \qquad
L_3 [f_9] = f_9 \, R_{\mu\nu\rho\sigma} \phi^\mu \phi^{\nu\rho} \phi^\sigma  \,, \label{Ncub4}
\eea
where $f_i$ are arbitrary functions of $\phi$ and $X$.
Only one of the two quadratic Lagrangians in (\ref{Nquad}) is independent, since it is possible to express one in terms of the other through integrations by parts: we worked with $L_2 [f_2]$.
Also the cubic Lagrangians (\ref{Ncub1} -- \ref{Ncub4}) are not all independent: we can obtain $L_3 [f_9]$ from $L_3 [f_6]$ and $L_3 [f_8]$ using integrations by parts, and $L_3 [f_8]$ from $L_3 [f_4]$, $L_3 [f_5]$ and $L_3 [f_3]$ using also the Bianchi identity.
Therefore, we are left with five cubic independent Lagrangians (\ref{Ncub1} -- \ref{Ncub3}). To keep  contact with Horndeski theory, without loss of generality it is useful to replace (\ref{Ncub1}) with the following expression
\begin{equation}
L_3 [f_3] = f_3 \, G_{\mu\nu}\,\phi^{\mu \nu} \,,
\label{GG}
\end{equation}
that we studied in the main text.

In this Appendix we concentrate separately on the four remaining cubic non-minimally coupled Lagrangians  (\ref{Ncub2} -- \ref{Ncub3}).
What characterises these Lagrangians in comparison with (\ref{GG}) is that they all feature time (and space) derivatives of the extrinsic curvature.
This indicates the possible presence of additional \Ost modes, this time coming from the metric sector of the theory,
unless there are suited extra primary constraints that remove them.

The covariant 3+1 decomposition of (\ref{Ncub2} -- \ref{Ncub3}) shows that the only components of the extrinsic curvature that acquire time derivatives are the scalar ones:
\beq
E \equiv \frac{\hat{A}^i \hat{A}^j}{\hat{A}^2} \, K_{ij} \,, \qquad F \equiv \hat{P}^{ij} K_{ij} \,.
\eeq
Their covariant velocities appear in the form
\beq
V_E \equiv n^{\mu} \nabla_{\mu} E \,, \qquad V_F \equiv n^{\mu} \nabla_{\mu} F \,,
\eeq
in analogy to what we encountered in section \ref{secADM}.
Therefore, applying the same kind of field redefinition used for the scalar field (\ref{lagmult}),
Lagrangians (\ref{Ncub2} -- \ref{Ncub3}) generally propagate two more \Ost modes, $E$ and $F$, in addition to $A_*$.
To avoid their propagation, we need two more primary constraints.

Defining the conjugate momenta associated to the new fields
\beq
 \pi_E \equiv 
 \frac{\delta {L}}{\delta V_E} \,, \qquad
 \pi_F \equiv 
 \frac{\delta {L}}{\delta V_F} \,,
 \eeq
for the set of Lagrangians (\ref{Ncub2} -- \ref{Ncub3}) we obtain
\beq
\pi_* =  \alpha \, V_E + \beta \, V_F + \dots \,,  \qquad
\pi_E =  \alpha \, V_* \,, \qquad
\pi_F =  \beta \, V_* \,, \label{piH}
\eeq
where
\bea
\alpha = - 2 f_4 - X f_6 + A_*^2 \left( 2 f_5 + X f_7 \right) \,, \qquad
\beta =  - 2 f_4 + A_*^2 \left( 2 f_5 + f_6 \right) -A_*^4 f_7 \,, \label{alphabeta}
\eea
and the dots in $\pi_*$ represent non relevant terms.
From the form of the momenta (\ref{piH}), it is clear that a total of three primary constraints can only be obtained in the trivial way, i.e. 
\beq
\pi_* \approx 0\,, \qquad \pi_E \approx 0 \,, \qquad \pi_F \approx 0 \,.
\eeq
Hence $\alpha = \beta = 0$ and, due to the Lorentz invariance of $f_i$, relations (\ref{alphabeta}) give
\beq
f_4=f_5=f_6=f_7=0 \,.
\eeq

\section{Tensorial structure implied by the degeneracy conditions}
\label{Tensorialstructure}

\no First, we show the equivalence between the relations (\ref{degeneracy_C2_intermsofC}) and (\ref{C2L}) for quadratic theories. It is simple to show that \eqref{degeneracy_C2_intermsofC} is equivalent to
\bea\label{MC2}
M^{\alpha\beta}_{\rho\sigma} \, C_\2^{\rho\sigma}(L)=0 \quad \text{with} \quad C_\2^{\rho\sigma}(L)\equiv C_\2^{\mu\nu\rho\sigma}L_{\mu\nu} 
\eea
and
\bea\label{M}
 M^{\alpha\beta}_{\rho\sigma} \; \equiv \; -\An \, g^{\alpha}_{(\rho} g^{\beta}_{\sigma)} \, + \, 2 n_{(\rho}g^{(\alpha}_{\sigma)} A^{\beta)} \, . 
\eea
Indeed, decomposing \eqref{MC2} in the directions $n_\alpha n_\beta$, $h_\alpha^i h_\beta^j$ and $n_\alpha h_\beta^i$ leads to the equations \eqref{degeneracy_C2_intermsofC}. As a consequence, $ C_\2^{\rho\sigma}(L)$ is necessarily
in the kernel of $M$ viewed as an operator acting on symmetric 4 dimensional matrices. A matrix $V^{\mu\nu}$ is in 
 the kernel of $M$ when
 \bea\label{kerM}
 M^{\alpha\beta}_{\mu\nu} \, V^{\mu\nu}=0  & \Longleftrightarrow  & \An V^{\alpha\beta} = 2 \, n_\mu V^{\mu (\alpha} A^{\beta)} \\
 & \Longleftrightarrow  & \exists \, V^\mu \, s.t. \, V^{\alpha\beta} = V^{(\alpha} A^{\beta)} \, \text{with} \, V_\mu n^\mu = 0 \, . 
 \eea
Furthermore, the only available vector $V^\mu$ in the theory which is orthogonal to $n_\mu$ is in the direction $\hat{A}^\mu$. Hence,  there exists a scalar $\m$ such that
\bea
C_{\2}^{\mu\nu\rho\sigma} L_{\rho\sigma}  = 2 \m A^{(\mu} \hat{A}^{\nu)}   \,, 
\eea
which is the relation (\ref{C2L}).

\smallskip

The generalization to cubic theories is rather immediate. Let us show that (\ref{C3L}) and (\ref{degeneracy_C3_intermsofC}) are equivalent. Following the same strategy as previously, we first
show that \eqref{degeneracy_C3_intermsofC} is equivalent to
\bea\label{MC}
M^{\alpha\beta}_{\rho\sigma} \, C_\3^{\mu\nu\rho\sigma}(L)=0 \quad \text{with} \quad C_\3^{\mu\nu\rho\sigma}(L)\equiv C_\3^{\mu\nu\rho\sigma\gamma\delta}L_{\gamma\delta} \, ,
\eea
with $M$ defined as in the quadratic case by (\ref{M}).

Now, both $M$ and $C_\3(L)$ can be viewed as operators acting on symmetric 4 dimensional matrices. Thus, 
 \eqref{MC} means that $C_\3(L)$ and $M$ are orthogonal, or equivalently that the image of $C_\3(L)$ lies in the kernel 
 of $M$. To go further, we recall that the kernel of $M$ is defined by (\ref{kerM}).
 The vector space orthogonal to $n^\mu$ is three dimensional and a basis is given by $h_\mu^\gamma$ where $\gamma$ labels the elements
  of the basis (only 3 out of the 4 components of $h_\mu^\gamma$ are independent).
 Thus, if we use the notation $V^{\mu\nu}_\gamma$ for a basis of $\text{Ker}(M)$ where $\gamma$ labels the elements of the basis, then
 $V^{\mu\nu}_\gamma = h^{(\mu}_\gamma A^{\nu)}$ which  is clearly of the form  \eqref{kerM}. 
  Hence, due to symmetries, $C_\3(L)$  can we written as
 \bea
 C_\3^{\mu\nu\rho\sigma}(L) =   m^{\gamma\delta} \, V^{\mu\nu}_\gamma V^{\rho\sigma}_\delta
 \eea
 where $m^{\gamma\delta}$ is a symmetric matrix. Due to the covariance, the symmetric matrix $m$ is necessarily of 
 the form $m^{\gamma\delta}= 4 \m_1 g^{\gamma\delta} + 4\m_2 A^\gamma A^\delta$
 where $\m_1$ and $\m_2$ are scalars. Notice that there is no components of the form $A^{(\gamma} n^{\delta)}$ nor 
 of the form $n^\gamma n^\delta$ in $m^{\gamma\delta}$ because $V^{\mu\nu}_\gamma n^\gamma=0$.
 As a conclusion, \eqref{MC} is true if and only if there exist scalars $\m_1$ and $\m_2$ such that:
  \bea\label{CL}
 C_\3^{\mu\nu\rho\sigma}(L) = 4 \m_1 \,  A^{(\mu} h^{\nu)(\rho} A^{\sigma)} + 4\m_2 \,  A^{(\mu}\tA^{\nu)} A^{(\rho}\tA^{\sigma)} \,  .
 \eea 

\section{Quadratic theories}
\label{app_quadratic}

\no We review the quadratic theories proposed in \cite{LN1} and further classified in \cite{Crisostomi:2016czh} and \cite{Achour:2016rkg}.

\subsection{Minimally coupled theories}

\no $\blacktriangleright$ {\bf $^2$M-I}: Three free functions $a_1, \, a_2$, and $a_3$, together with
\bea
a_4 = - 2\,\frac{a_1}{X} \, , \qquad a_5 = \frac{4 a_1 \left( a_1 + 2 a_2 \right) - 4 a_1 a_3 X + 3 a_3^2 X^2}{4 \left(a_1 + 3 a_2\right) X^2}.
\eea

We assume $a_2 \neq - a_1/3$. This case includes beyond Horndeski theory.
The corresponding null eigenvector is given by
\bea
v_1 &=& - \frac{X (2a_2+a_3X)}
{A_* \left(2 a_1 \left(A_*^2+2X\right) + 2 a_2 \left(2 A_*^2 + 5X\right) - a_3 X\left(A_*^2+X\right)\right)} \,, \\
v_2 &=& \frac{-2 a_1 - 4 a_2 + a_3 X}
{A_* \left(2 a_1 \left(A_*^2+2X\right) + 2 a_2 \left(2 A_*^2 + 5X\right) - a_3 X\left(A_*^2+X\right)\right)} \,.
\eea
This class was called M-I in \cite{Crisostomi:2016czh} and IIIa in \cite{Achour:2016rkg}.

\hspace{1cm}\\
$\blacktriangleright$  {\bf $^2$M-II}: Three free functions $a_1$, $a_4$, $a_5$ and
\bea
a_2 = - \frac{a_1}{3} \,, \qquad a_3 = \frac{2\,a_1}{3\,X} \,.
\eea
The corresponding null eigenvector is given by
\bea
v_1= \frac{X}{A_*(A_*^2+X)} \,, \qquad v_2= - \frac{1}{A_*(A_*^2+X)} \, .
\eea
This class was called M-II in \cite{Crisostomi:2016czh} and IIIb in \cite{Achour:2016rkg}.

\hspace{1cm}\\
$\blacktriangleright$  {\bf $^2$M-III}: Four free functions $a_2,a_3,a_4, a_5$ and the unique condition
\bea
a_1=0 \,.
\eea
The eigenvector is given by
\bea
v_1= -\frac{X}{2A_*(A_*^2+X)} \,, \qquad v_2= \frac{-2A_*^2+X}{2A_*(A_*^2+X)^2} \, .
\eea

This class was called M-III in \cite{Crisostomi:2016czh} and IIIc in \cite{Achour:2016rkg}.

For minimally coupled quadratic theories, the vector components of $\pi_{ij}$ (i.e. $\tA_i \hat{P}_{jk} \pi^{ij}$) are proportional to $a_1$, therefore, as noticed in \cite{deRham:2016wji}, this class propagates only one scalar dof.

\subsection{Non-minimally coupled theories}

\no $\blacktriangleright$ {\bf $^2$N-I}: Three free functions $f_2, a_1$ and $a_3$. The conditions are
\bea
a_2&=& - a_1 \neq -\frac{ f_2}{X} \, , \\
\label{A4}
a_4&=& \frac{1}{8(f_2 - a_1 X)^2}\left[ 4 f_2 \left(3 (a_1 - 2 f_{2X})^2 - 2 a_3 f_2\right) -a_3 X^2 (16 a_1 f_{2X} + 
a_3 f_2)  \right. \nb \\
&& \left. \qquad\qquad\qquad + 4 X \left(3 a_1 a_3 f_2 + 16 a_1^2 f_{2X} - 16 a_1 f_{2X}^2 - 4 a_1^3 + 2 a_3 f_2 f_{2X}\right) \right] \,,   \\
\label{A5}
a_5 &=& 
\frac{1}{8 (f_2 - a_1 X)^2}(2 a_1 - a_3 X - 4 f_{2X}) \left[a_1 (2 a_1 + 3 a_3 X - 4 f_{2X}) - 4 a_3 f_2\right] \,. 
\eea
The combination of Horndeski and beyond Horndeski theories is included in this class.
The corresponding null eigenvector is given by
\bea
v_1 & = & D \,  A_* (a_1 X - f_2) (2a_1 - a_3 X - 4 f_{2X}) \, , \\
v_2 & = & D \, A_* \left(a_1 (2 a_1 + a_3 X - 4 f_{2X}) - 2 a_3 f_2\right) \, ,
\eea
with
\bea
D^{-1} & \equiv & a_1\left(A_*^2 \left(-a_3 X^2 + 2f_2 + 12 f_{2X} X\right) + A_*^4 (4f_{2X} - a_3 X) - 6 f_2 X+8 f_{2X} X^2\right) \nb \\
&& -2 a_1^2 \left(3A_*^2 X + A_*^4\right) + f_2\left(\left(A_*^2 + X\right) \left(a_3\left(2 A_*^2 + X\right) - 4 f_{2X}\right) + 4f_2\right)\, . \nb
\eea

This class was called N-I in \cite{Crisostomi:2016czh} and Ia in \cite{Achour:2016rkg}.

\hspace{1cm}\\
$\blacktriangleright$  {\bf $^2$N-II}: Three free functions $f_2, a_4, a_5$ and
\bea
a_2= - a_1 = - \frac{f_2}{X} \,, \qquad a_3 =  \frac{2\left(f_2 - 2 X f_{2X}\right)}{X^2}
\eea
The corresponding null eigenvector is given by
\bea
v_1 = 0 \,, \qquad v_2 = -\frac{A_* }{\left(A_*^2 + X\right)^2} \, .
\eea

This class was called N-II in \cite{Crisostomi:2016czh} and Ib in \cite{Achour:2016rkg}.

For non-minimally coupled quadratic theories, the vector components of $\pi_{ij}$ are instead proportional to $f_2 - X a_1$, hence also here there are not tensor modes \cite{deRham:2016wji}.

\hspace{1cm}\\
$\blacktriangleright$  {\bf $^2$N-III}: Three free functions $f_2, a_1$ and $a_2$. The conditions are
\bea
a_1 + a_2 \neq 0\,, \qquad \text{and} \qquad a_1 \neq \frac{f_2}{X},
\eea
\bea
a_3 &=& \frac{4 f_{2X} (a_1 + 3 a_2)}{f_2} - \frac{2 (a_1 + 4 a_2 - 2 f_{2X})}{X} - \frac{4 f_2}{X^2} \,, \\
a_4 &=& \frac{2 f_2}{X^2} + \frac{8 f_{2X}^2}{f_2} - \frac{2 (a_1 + 2 f_{2X})}{X} \,, \\
a_5 &=& \frac{2}{f_2^2 X^3} \left[4 f_2^3 + f_2^2 X (3 a_1 + 8 a_2 - 12 f_{2X}) \right. \nb \\
&& \left. \qquad\quad + 8 f_2\, f_{2X} X^2 (f_{2X} - a_1 - 3 a_2)+6 f_{2X}^2 X^3 (a_1 + 3 a_2)\right] \,.
\eea
The corresponding null eigenvector is given by
\bea
v_1 &=& \frac{X (f_2 - 2 f_{2X} X)}{A_*\left(2 A_*^2 (f_2 - f_{2X} X) + X (3f_2 - 2 f_{2X} X)\right)} \,, \\
v_2 &=& \frac{2 f_{2X} X - 2 f_2} {A_* \left(2A_*^2 (f_2 - f_{2X} X) + X (3f_2 - 2 f_{2X} X)\right)} \, .
\eea
This class was called N-III (i) in \cite{Crisostomi:2016czh} and IIa in \cite{Achour:2016rkg}.

\hspace{1cm}\\
$\blacktriangleright$  {\bf $^2$N-IV}: Three free functions $f_2,\, a_2$ and $a_3$. The conditions are
\bea
a_1 + a_2 \neq 0
\eea
\bea
a_1&=&\frac{f_2}{X} \, , \\
a_4 &=&\frac{8 f_{2X}^2}{f_2} - \frac{4 f_{2X}}{X} \,, \\
a_5 &=& \frac{1}{4 f_2 X^3 (f_2 + a_2 X)} \left[ f_2 a_3^2 X^4 - 4 f_2^3 - 8 f_2^2 X (a_2 - 2 f_{2X})
 \right. \nb \\
&& \left. - 4 f_2 X^2 \left(4 f_{2X} \left(f_{2X} - 2 a_2\right) + a_3 f_2\right) + 8 f_{2X} X^3 (a_3 f_2 - 4 a_2 f_{2X}) \right] \,.
\eea
The corresponding null eigenvector is given by
\bea
v_1 &=&  - 2\, E \,  X (a_2 X + f_2) (f_2 - 2 f_{2X} X)  \, ,\\
v_2 &=& E\frac{4 X (a_2 X + f_2) (f_2 - 2 f_{2X} X) -A_*^2 \left(f_2 X (4a_2 + a_3 X - 4 f_{2X}) - 8 a_2 f_{2X} X^2 + 2 f_2^2\right)}{A_*^2 +X} \, ,
\eea
with
\bea
E^{-1}&\equiv& A_*^3 \left(f_2 X (4a_2 + a_3 X - 4 f_{2X}) - 8 a_2 f_{2X} X^2 + 2 f_2^2\right) \nb \\ 
&& + A_* X^2(2 a_2 (f_2 - 4 f_{2X} X) + a_3 f_2 X - 4 f_2 f_{2X}) \,. \\
\eea

This class was called N-III (ii) in \cite{Crisostomi:2016czh} and IIb in \cite{Achour:2016rkg}.

Due to the condition $f_2 - X a_1 =0$, only one scalar dof is present in this class \cite{deRham:2016wji}.

 \section{Identifying cubic theories}
 \label{app_cubic}
\no In this Appendix we solve in details the conditions (\ref{C3L}) for purely cubic theories.
Let us first note that, since $\V_{ij}$ lies in the hyper-surface orthogonal to $n^\mu$, it can be decomposed as follows
\bea
\label{V}
\V_{ij}= v_1 h_{ij} + v_2 \tA_i \tA_ j \, ,
\eea
where $v_1$ and $v_2$ are scalar quantities.
The tensor $L_{\mu\nu}$, introduced in \eqref{L}, can thus be written  as 
\bea\label{def of L}
L_{\mu\nu}  \equiv  \lambda_{\mu\nu} + \Lambda_{\mu\nu}^{ij} \V_{ij} 
&=&\lambda_{\mu\nu} + v_1 \Lambda_{\mu\nu}^{ij} h_{ij}+v_2\Lambda_{\mu\nu}^{ij} \tA_i \tA_ j  \,,
\cr
 &=& \l_1 \, n_\mu n_\nu + 2\, \l_2 \, n_{(\mu} A_{\nu)} - \An (v_1 g_{\mu\nu} + v_2 A_\mu A_\nu) \,,
\eea
with
\bea
\l_1 \equiv 1+\An v_1 + \An(2X + \An^2) v_2  \qquad \text{and} \qquad
\l_2 \equiv  v_1 + X v_2 \, .
\eea

\subsection{Minimally coupled theories}
 \no A long but straightforward calculation shows
 that the tensorial relation \eqref{C3L}  leads to the following 12 equations:
\bea
&&\l_1 \b_2 = 0 \,, \quad \l_1 \b_3 = 0 \, , \quad \l_1 b_6 = 0 \, ,\\
&&(\l_1 \An + X \l_2)\b_5 + 2 \l_2 \b_2  = 0 \, , \quad (\l_1  \An + X \l_2)\b_7 + 3 \l_2 \b_3  =  0 \, , \\  
&&\l_1 \b_7 = 12(m_1 + \An^2 m_2) \, , \quad (\l_1 \An + X \l_2)\b_8 + \l_2 (2\b_6 + \b_7)=  12 \An m_2 \label{eq for mus} \, , \\
&& \l_1 (- \b_9 + 3 \An^2 \b_{10}) + 2 \l_2 \An (3X \b_{10} +  \b_8 + \b_9) \nb \\
&&-\An[ (v_1 + X v_2) ({2}\b_8 + 3X\b_{10}) + (4v_1 + X v_2) \b_9 + {2} v_2 \b_6 +  v_2 \b_7] = 12 m_2 \, ,\\
&&\l_1(- \b_4 +  \An^2 \b_9) + \l_2 \An ({2} \b_9 X + {2} \b_4 + \b_5) \nb \\
&&- \An [ \b_4 (4v_1 + Xv_2) + (\b_5 + X \b_9) (v_1 + X v_2) + (\b_6 v_1 + \b_2 v_2))] = 0 \, ,\\
&& \l_1 (- \b_5 + \b_8 \An^2) + 2\l_2 \An ( \b_8 X +  \b_5 + \b_7) \nb \\
&&-\An [ \b_5 (4v_1+Xv_2) + 3 \b_3 v_2 + \b_7 (3v_1 + 2X v_2) +  \b_8 X(v_1+Xv_2)] = 12 m_1 \, ,\\
&& \l_1 (-3\b_1 +  \b_4 \An^2) + 2 \l_2 \An (\b_4 X + 3\b_1) \nb \\
&&-\An [3\b_1 (4v_1 + v_2 X) + {2} \b_2 v_1 +  \b_4 X (v_1 + X v_2)] = 0 \, ,\\
&& \l_1 (-\b_2 + \b_6 \An^2) + 2 \l_2 \An ( \b_6 X +  \b_2) \nb \\
&& - \An [ \b_2 (4v_1+Xv_2) + 3 \b_3 v_1 +  \b_6 X (v_1 + X v_2)] = 0 \, . 
\eea
The two equations in \eqref{eq for mus} enable us to solve for  $\m_1$ and $\m_2$, yielding
\bea
\m_1 & = & \frac{1}{12} \left[ (\l_1 - \An \l_2) \b_7 - \An (\l_1 \An + X \l_2) \b_8 - 2\An \l_2 \b_6\right]\, ,\\
\m_2 & = & \frac{1}{12 \An} \left[(\l_1 \An + X \l_2)\b_8 + \l_2 (2\b_6 + \b_7) \right] \, .
\eea
Hence, if we replace these expressions in the previous system, we end up with 10 equations for the 10 unknown $\b_i$. These 10 equations can be written in a matrix form as follows:
\bea\label{Matrix notation}
\left(
\begin{array}{cccccc|cccc}
&&&&&&&&& \\
&&&&&&&&& \\
&&&& &&& && \\
&&{\bf A }&&&&&&{\bf 0} & \\
&&&&&&&&& \\
&&&&&&&&& \\
\hline
&&&&&&&&& \\
&& {\bf C}&&&&& & {\bf B} & \\
&&&&&&&&& \\
&&&&&&&&& 
\end{array}
\right)
\left(
\begin{array}{c}
\b_2 \\
\b_3 \\
\b_6 \\
\b_5 \\
\b_7 \\
\b_{1} \\
\hline
\b_4 \\
\b_8 \\
\b_9 \\
\b_{10}
\end{array}
\right) = 0 \,,
\eea
where 
\bea
{\bf A} & \equiv &  
\left(
\begin{array}{cccccc}
\l_1 & 0 & 0  & 0 & 0 & 0   \\
0   & \l_1 & 0  & 0 & 0 & 0  \\
0   & 0 & \l_1 & 0 & 0 & 0   \\
2\l_2    & 0 & 0 & T & 0 & 0  \\
  0   & 3\l_2 &  0 & 0 & T & 0  \\
  S & - 3 \An v_1 & \An T & 0 & 0 & 0 
  \end{array}
 \right) \;\; , \;\;
 {\bf B}  \, \equiv \, 
\left(
\begin{array}{cccc}
  \An T & 0 & 0 & 0 \\
 0 & 2\An T & 0 & 0 \\
  S & 0 & \An T  & 0 \\
 0 & -T & S \An & 3\An^2 T   
\end{array}
\right) \, ,
 \\
{\bf C} & \equiv & 
\left(
\begin{array}{cccccc} -2\An v_1 & 0 & 0 &  0 & 0 & 3S  \\
   0  & -3\An v_2 & 2\An \l_2 & S &  - \l_1 + X \An v_2  & 0  \\
  -\An v_2    & 0 & -\An v_1 & 0 & 0 & 0  \\
  0   & 0 & -2(\An^2 v_2 + \l_2) & 0 &  -(\An^2 v_2 + \l_2) & 0
\end{array}
\right)  \,,  \nb
\eea
and $\bf 0$ denotes a $6\times 4$ matrix of zeros. 
We have also introduced  the notation $T\equiv\l_1 \An + X \l_2$ and $S\equiv-\l_1 + \l_2 \An - 3 v_1 \An$. 

The resolution of the system depends on the rank of the matrices $\bf A$ and $\bf B$. To solve the system, it is useful to separate the vector in \eqref{Matrix notation} into two pieces
\bea
{\bf \b}_{+}=(\b_2, \b_3, \b_6, \b_5, \b_7 , \b_1) \quad \text{and} \quad {\bf \b}_-=(\b_4,\b_8,\b_9,\b_{10}) \, .
\eea
Hence, we solve successively the following two matrix equations
\bea
{\bf A} {\bf \b}_+ = 0  \qquad \text{and} \qquad {\bf C} {\bf \b}_+ + {\bf B}{\bf \b}_-=0 \, . 
\eea
We can distinguish several cases, depending on  whether $\l_1$ or $T$ vanish. 

\renewcommand{\theenumi}{$^3$M-\Roman{enumi})}

\hspace{1cm}\\
$\blacktriangleright$ $\l_1 = 0$ and $T \neq 0$ \\

In that case $v_2$ is related to $v_1$ by
\bea
v_2 = -\frac{1+A_* v_1}{A_* (A_*^2+2X)} \, .
\eea
The matrix $\bf A$ is highly degenerate with rank=$3$ and ${\bf A} {\bf \b}_+ = 0$ produces 3 conditions. 
Two of them give
\bea
\b_5 = - \frac{2}{X} \b_2 \, , \qquad \b_7 = - \frac{3}{X} \b_3 \,,
\eea
and the third one is
\bea
\An \left( \An^2(2\b_2 + 3\b_3 - X \b_6) + X(5 \b_2 + 6 \b_3 - X \b_6)  \right) v_1 \, = \, - X(\b_2 + X \b_6) \, . \label{v1case2}
\eea
Equation (\ref{v1case2}) plus the four remaining  equations ${\bf C} {\bf \b}_++ {\bf B} {\bf \b}_-=0$ are solved into three sectors.
\begin{enumerate}
\item $9\b_1+ 2 \b_2 \neq 0$:
\bea
\b_6 & = & \frac{9 \b_1 \b_3 + 3 \b_4 X(\b_2 + \b_3) - 2 \b_2^2}
{X (9\b_1 + 2 \b_2)} \,,\nb \\
\b_8 & = & \frac{9 \b_1 \b_3 - 6 \b_4 X(\b_2 + \b_3) + 6 \b_2 \b_3 + 4 \b_2^2}
{X^2 (9 \b_1 + 2 \b_2)} \,,\nb\\
\b_9 & = & \frac{1}{X^2 (9 \b_1 + 2 \b_2)^2} \Big[3 \b_4^2 X^2 (9 \b_1 + 3 \b_2 + \b_3) - 2 \b_4 X \left(9 \b_1 (\b_2 - \b_3) + 4 \b_2^2\right) \nb \\
&& + 24 \b_1 \b_2^2 + 54 \b_1^2 \b_2 + 27 \b_1^2 \b_3 + 4 \b_2^3\Big] \,,\nb \\
\b_{10} & = & \frac{1}{X^3 (9 \b_1 + 2\b_2)^3} \Big[3 \b_4^3 X^3 (9 \b_1 + 3 \b_2 + \b_3) - 6 \b_2 \b_4^2 X^2(9 \b_1 + 3 \b_2 + \b_3) \nb \\
&& + 2 \b_4 X \left(81 \b_1^2 (\b_2 + \b_3) + 18 \b_1 \b_2 (3 \b_2 + 2 \b_3) + 2 \b_2^2 (5 \b_2 + 3 \b_3)\right) \nb \\
&& - 2\left(54 \b_1^2 \b_2 (\b_2 + 2 \b_3) + 4 \b_1 \b_2^2 (7 \b_2 + 9 \b_3) + 81 \b_1^3 \b_3 + 4 \b_2^3(\b_2 + \b_3)\right) \Big] \,,\nb
\eea
and therefore
\bea
v_1 = -\frac{X (3 \b_1 + \b_4 X)}{A_* \left(A_*^2 (6 \b_1 + 2 \b_2 - \b_4 X) + X (15 \b_1 + 4 \b_2 - \b_4 X)\right)}
\eea
As a conclusion, we end up with four free parameters $\b_1,\b_2,\b_3$ and $\b_4$. 
\item $9\b_1 + 2\b_2 = 0$ and $9 \b_1 - 2 \b_3 \neq 0$: 
\bea
\b_2&=&-\frac{9}{2} \b_1 \, , \qquad
\b_4 = -\frac{3}{X} \b_1 \, , \qquad
\b_8=\frac{3\b_3-2\b_6 X}{X^2} \, , \nb \\
\b_9 &=& \frac{9 \b_1 (\b_3 + 2 \b_6 X) - 81 \b_1^2 - 2 \b_6^2 X^2}{3 X^2 (9 \b_1 - 2 \b_3)} \, , \nb \\
\b_{10} &=& \Big[18 \b_6 X \left(-12 \b_1 \b_3 + 27 \b_1^2 + 2 \b_3^2\right) - 36 \b_3\left(-8 \b_1 \b_3 + 18 \b_1^2 + \b_3^2\right) \nb \\ 
&& - 12 \b_3 \b_6^2 X^2 + 4 \b_6^3 X^3\Big]
\Big[9 X^3 (9 \b_1 - 2 \b_3)^2\Big]^{-1} \,, \nb
\eea
and
\bea
v_1 = \frac{X (2 \b_6 X - 9 \b_1)}{A_*\left[ 2 A_*^2(9 \b_1 - 3 \b_3 + \b_6 X) + X (45 \b_1 - 12 \b_3 + 2 \b_6 X) \right]}
\eea
The three parameters $\b_1,\b_3,\b_6$ are free.
\item $9\b_1 + 2\b_2 = 0$ and $9 \b_1 - 2 \b_3 = 0$: 
\bea
\b_2 &=&-\frac{9 \b_1}{2} \, , \qquad \b_3 = \frac{9 \b_1}{2} \,, \qquad \b_4 = - \frac{3 \b_1}{X} \,, \nb \\
\b_6 &=& \frac{9 \b_1}{2X} \,, \qquad \b_8  =  \frac{9 \b_1}{2X^2} \,, \qquad \b_9 = - \frac{3\, \b_1}{2X^2} \, , \qquad \b_{10} = - \frac{\b_1}{X^3} \, . \nb
\eea
We obtain that $\b_1$ is free whereas there is no constraint on $v_1$. 
\end{enumerate}

\hspace{1cm}\\
$\blacktriangleright$ $\l_1= 0$ and $T=0$:  $^3$M-IV \\

This case is characterized by the fact that
$v_1$ and $v_2$ are totally fixed by
 \beq
v_1=\frac{X}{\An (X+\An^2)}\,, \qquad v_2= -\frac{1}{\An (X+\An^2)} \, .
\eeq
Furthermore, ${\bf A}{\bf \b}_+=0$ fixes 
\bea
\b_3=-\b_2 \,.
\eea
The four remaining equations give
\bea
&& \b_2=-\frac92 \b_1 \,, \qquad  \b_6=-3\b_4-\frac{9}{2X}\b_1 \,, \nb \\[2ex]
&&\b_7=-3\b_5 + \frac{27}{2X}\b_1 \,, \qquad \b_9= \frac{3 \b_1 - 2 X (2 \b_4 + \b_5)}{2 X^2} \, , \nb
\eea
and $\b_1,\b_4,\b_5,\b_8,\b_{10}$ are free.

\hspace{1cm}\\
$\blacktriangleright$ $\l_1 \neq 0$ and $T \neq 0$: $^3$M-V \\

In that case, $\bf B$ is invertible and 
$\bf A$ reaches its maximal rank$=5$. Hence, from ${\bf A} {\bf \b}_+ = 0$  we get
\bea
\b_2=\b_3=\b_5=\b_6=\b_7=0 \,.
\eea 
The four remaining equations  ${\bf C} {\bf \b}_+ + {\bf B}{\bf \b}_-=0$ give
\bea
\b_8=0 \,,
\eea
together with three equations. If $\b_1=0$ all the other functions must be zero, therefore we assume $\b_1 \neq 0$ and
we obtain
\bea
\b_9 = \frac{\b_4^2}{3\b_1} \,,  \qquad \, \b_{10}=\frac{\b_4^3}{27 \b_1^2} \, ,
\eea
plus one relation between $v_1$ and $v_2$:
\bea
A_* \b_4 ( \An + \left(A_*^2 + X \right)  v_1 + \left(A_*^2 + X \right)^2 v_2)= 3 \b_1\left(1+ 3\An v_1 + A_*  \left(A_*^2 + X \right) v_2 \right) \ .
\eea
As a conclusion, only two parameters, $\b_1$ and $\b_{4}$, are free. One of the two components $v_1$ or $v_2$ of the eigenvector is also a free parameter.

This class possesses two more primary constraints of the form (\ref{primaryvec}), hence there is only one scalar dof.

\hspace{1cm}\\
$\blacktriangleright$ $\l_1 \neq 0$ and $T=0$ \\

$v_1$ is fixed by 
\bea
v_1 = - \frac{\An}{X+ \An^2} - (X+\An^2) v_2 \, .
\eea
Solving ${\bf A}{\bf \b}_+=0$ leads to 
\bea
\b_2=\b_3=\b_6=0 \,.
\eea
Furthermore, the equations ${\bf C} {\bf \b}_+ + {\bf B}{\bf \b}_-=0$ give two branches:

\begin{enumerate} 
\addtocounter{enumi}{5}

\item $\b_7=0$

and the component $v_2$ is fixed to
\bea
v_2= \frac{X -2 A_*^2}{2 A_* \left(A_*^2 + X\right)^2} \,.
\eea
The components $\b_1$, $\b_4$, $\b_5$, $\b_8$, $\b_9$ and $\b_{10}$ are free.

Also in this class, tensor modes are eliminated by the primary constraints (\ref{primaryvec}), so only one scalar dof is left.

\item $\b_1=\b_4=0 \,, \qquad \b_9=- \b_5/X$

and the component $v_2$ is fixed by
\bea
2 A_* \b_5 \left(A_*^2 + X\right)^2\, v_2= X (\b_5 + \b_7) - 2 A_*^2 \b_5 \, .
\eea
The components $\b_5$, $\b_7$, $\b_8$ and $\b_{10}$ are free.
\end{enumerate}

\subsection{Non-minimally coupled theories}

\no The resolution follows the same strategy as in the minimally coupled case.
First, we write the generalised conditions in a form analogous to \eqref{Matrix notation} 
\bea
\left(
\begin{array}{cc}
{\bf A} & {\bf 0} \\
{\bf C} & {\bf B}
\end{array}
\right) 
\left(
\begin{array}{c}
{\bf \tilde \b}_+ \\
{\bf \tilde \b}_-
\end{array}
\right)= \frac{X\, f_{3X}}{A_*}\, {\bf \Sigma} \,, \label{non-min}
\eea
where ${\bf \Sigma}$ is a  matrix given by
\bea
{\bf \Sigma}=
\left(
\begin{array}{c}
- \l_2 \\
2 \l_2/3 \\
v_2  \\
2\An v_2 \\
-2\An v_2 \\
 \l_2 + \An^2 v_2 \\
- (\l_2 + \An^2 v_2) \\
0\\
v_2 \\
0
\end{array}
\right) \,.
\eea
Hence, the solution  for ${\bf \tilde \b}=({\bf \tilde \b}_+,{\bf \tilde \b}_-)$ is the sum of the general
solution of the homogeneous equation (with $f_{3}=0$) and a particular solution. 
Again, we solve them according to whether $\l_1$ and $T$ vanish or not.

When $\l_1=0$ necessarily $v_1=v_2=0$, which would imply in turn that $\l_1=1$. This is an inconsistency, hence
there is no solution when $\l_1=0$.

\hspace{1cm}\\
$\blacktriangleright$ $\l_1 \neq 0$ and $T \neq 0$: $^3$N-I \\

We need to assume $\b_1 \neq 0$ otherwise $T = 0$, and we end up in the next class of theories.
\bea
&& \b_2=  -3\,\b_1 \, , \qquad \b_3 = 2\,\b_1 \, , \qquad \b_6 = -\b_4 \, , \nb \\[2ex]
&& \b_5 = \frac{2 (f_{3X} - 3 \b_1)^2 - 2 \b_4 f_{3X} X}{3 \b_1 X} \, ,\qquad \b_7 = \frac{2 \b_4 f_{3X} X - 2 (f_{3X} - 3 \b_1)^2}{3 \b_1 X} \, , \nb \\
&& \b_8 = \frac{2 (3 \b_1 + \b_4 X - f_{3X})\left((f_{3X} - 3 \b_1)^2 - \b_4 f_{3X} X\right)}
{9 \b_1^2 X^2} \,, \nb \\
&& \b_9 = \frac{2 \b_4 (3 \b_1 + \b_4 X - f_{3X})}
{3 \b_1 X} \, ,\qquad \b_{10} = \frac{2 \b_4 (3 \b_1 + \b_4 X - f_{3X})^2}
{9 \b_1^2 X^2} \, ,  \nb
\eea
and $\b_1$, $\b_4$ and $f_3$ are free. Furthermore, the eigenvector is given by
\bea
v_1&=& \frac{A_* (3 \b_1 + \b_4 X - f_{3X})}
{A_*^2 (\b_4 X - 3 \b_1 + f_{3X}) + A_*^4 \b_4 + f_{3X} X} \, , \\[2ex]
v_2 &=& -\frac{A_* \b_4}
{A_*^2 (\b_4 X - 3 \b_1 + f_{3X}) + A_*^4 \b_4 + f_{3X} X} \, . 
\eea
This is the particular solution, now we need to find which homogeneous solution is compatible with it.
To ensure the full theory to be degenerate, the eigenvectors of the homogeneous and the particular solutions must coincide.
It is easy to show that it cannot be supplemented with any of the minimally coupled theories. 

\hspace{1cm}\\
$\blacktriangleright$ $\l_1 \neq 0$ and $T=0$: $^3$N-II \\

We obtain
\bea
&& \b_1 = \b_2= \b_3 = \b_5 = \b_7 = 0 \, , \nb \\[2ex]
&& \b_4 = -\b_6= \frac{f_{3X}}{X}  \, ,\qquad \b_9 = -\frac{2\,f_{3X} }{X^2} \, . \nb
\eea
Furthermore, the eigenvector is given by
\bea\label{NMclass2v}
v_1=0 \, , \qquad v_2= - \frac{\An}{(X+\An^2)^2} \, .
\eea
Now we study which homogeneous solution can be added to this particular one.
It is possible to add only $^3$M-VII where $\b_5+\b_7=0$.
Therefore, the full conditions for this class of theories are
\bea\label{NMclass2-2}
&& \b_1 = \b_2= \b_3 =0 \,, \qquad \b_7 = - \b_5  \, , \nb \\[2ex]
&& \b_4 = -\b_6= \frac{f_{3X}}{X}  \, ,\qquad \b_9 = -\frac{2\,f_{3X} + X \b_5 }{X^2} \, . \nb
\eea
and $\b_5,\, \b_8,\, \b_{10}$ and $f_3$ are free.

\end{document}